\documentclass[10pt,journal,compsoc]{IEEEtran}
\usepackage{amsmath,amssymb,amsfonts}
\usepackage{algorithmic}
\usepackage{graphicx}
\usepackage{textcomp}
\usepackage{xcolor}
\usepackage{comment}
\usepackage{fontawesome}
\usepackage{csquotes}
\usepackage{relsize,etoolbox}
\usepackage[rightmargin=0.5pt,leftmargin=0.5pt]{quoting}
\usepackage{booktabs} 

\usepackage{xspace}
\usepackage[normalem]{ulem}
\useunder{\uline}{\ul}{}
\usepackage{multirow}
\usepackage{lscape}
\usepackage{rotating}
\usepackage{supertabular}
\usepackage{longtable}
\usepackage{enumerate}
\usepackage{colortbl}
\usepackage{subcaption}
\usepackage{hhline}
\usepackage{makecell}
\usepackage{caption}
\usepackage{hyperref}
\usepackage{enumitem}
\usepackage{tcolorbox}
\usepackage{tabularx}
\usepackage{lipsum}
\usepackage{balance}
\usepackage{blindtext,graphicx}
\usepackage[absolute]{textpos}
\usepackage{fdsymbol}
\usepackage{url}

\usepackage{breakurl}

\newcommand{\todo}[1]{}
\renewcommand{\todo}[1]{{\color{red} TODO: {#1}}}

\ifCLASSINFOpdf
\else
\fi
\begin{document}
%
\title{Supporting Developers in Addressing Human-centric Issues in Mobile Apps}



%


\author{Hourieh~Khalajzadeh,
        Mojtaba~Shahin,
        Humphrey~O.~Obie,
        Pragya~Agrawal, and
        John~Grundy
\IEEEcompsocitemizethanks{\IEEEcompsocthanksitem Hourieh Khalajzadeh is with the School of Information Technology, Faculty of Science Engineering and Built Environment, Deakin University, Melbourne, Australia 3125.\protect\\
E-mail: hkhalajzadeh@deakin.edu.au

\IEEEcompsocthanksitem Mojtaba Shahin is with the School of Computing Technologies, RMIT University, Melbourne, Australia, 3000.\protect\\
E-mail: mojtaba.shahin@rmit.edu.au

\IEEEcompsocthanksitem Humphrey Obie is with the Department of Software Systems and Cybersecurity, Faculty of IT, Monash University, Melbourne, Australia, 3800.\protect\\
E-mail: humphrey.obie@monash.edu

\IEEEcompsocthanksitem Pragya Agrawal is with the Department of Software Systems and Cybersecurity, Faculty of IT, Monash University, Melbourne, Australia, 3800.\protect\\
E-mail: pagr0009@student.monash.edu

\IEEEcompsocthanksitem John Grundy is with the Department of Software Systems and Cybersecurity, Faculty of IT, Monash University, Melbourne, Australia, 3800.\protect\\
E-mail: john.grundy@monash.edu

\IEEEcompsocthanksitem Hourieh Khalajzadeh is the corresponding author.
}
\thanks{Manuscript accepted 23 Sept 2022.}
}

\markboth{IEEE Transactions on Software Engineering
,~Vol.XX, No.XX, September~2022}%
{Khalajzadeh \MakeLowercase{\textit{et al.}}: Bare Advanced Demo of IEEEtran.cls for IEEE Computer Society Journals}



\IEEEtitleabstractindextext{
\begin{abstract} 
Failure to consider the characteristics, limitations, and abilities of diverse end-users during mobile app development may lead to problems for end-users, such as accessibility and usability issues. We refer to this class of problems as \emph{human-centric issues}. Despite their importance, there is a limited understanding of the types of human-centric issues that are encountered by end-users and taken into account by the developers of mobile apps. 
In this paper, we examine what human-centric issues end-users report through Google App Store reviews, what human-centric issues are a topic of discussion for developers on GitHub, and whether end-users and developers discuss the same human-centric issues. We then investigate whether an automated tool might help detect such human-centric issues and whether developers would find such a tool useful. To do this, we conducted an empirical study by extracting and manually analysing a random sample of 1,200 app reviews and 1,200 issue comments from 12 diverse projects that exist on both Google App Store and GitHub. Our analysis led to a taxonomy of human-centric issues that characterises human-centric issues into three-high level categories: App Usage, Inclusiveness, and User Reaction. We then developed machine learning and deep learning models that are promising in automatically identifying and classifying human-centric issues from app reviews and developer discussions. A survey of mobile app developers shows that the automated detection of human-centric issues has practical applications. Guided by our findings, we highlight some implications and possible future work to further understand and better incorporate addressing human-centric issues into mobile app development.
\end{abstract}

\begin{IEEEkeywords}
Human-centric issues, GitHub repositories, Google Play Store, human aspects, machine learning, deep learning
\end{IEEEkeywords}}

\maketitle
%
\IEEEdisplaynontitleabstractindextext

\IEEEpeerreviewmaketitle

\section{Introduction} \label{sec:Introduction}
Even though software systems, including mobile applications (apps), are designed to fulfill the expectations of their diverse end-users, negligence of \textit{human-centric issues} during the software development process can lead to hard-to-deploy, hard-to-maintain, and hard-to-use software \cite{grundy2020towards,hartzel2003self,miller2015emotion,stock2008evaluation,wirtz2009age}. \textit{We define human-centric issues as the problems end-users might face when using mobile apps that stem from the lack of (proper) consideration of their specific/differing human characteristics, limitations, abilities, and personalities.}
Mobile app developers need to be aware and carefully investigate the human-centric issues~\cite{2007Kulyk}. However, they are not necessarily aware of, have not experienced, or do not understand and effectively communicate the implications of such issues. 
There is no evidence-based knowledge about how different types of human-centric issues are reported by the end-users and whether they are sufficiently being discussed and addressed during mobile apps development. This work aims to develop a comprehensive taxonomy of human-centric issues, and also to investigate whether an automated tool can help effectively detect such human-centric issues. We are ultimately interested in understanding how software/app developers perceive the usefulness of such a tool.

App reviews are a rich resource of the issues that end-users face when using an app. These include the human-centric issues that we are interested to learn more about, from the end-users' perspective. Additionally, developer discussions can be a major factor in deciding how a mobile app evolves, and include information beyond how an app works \cite{brunet2014developers,tsay2014let}. Online software repositories, e.g., GitHub, attract a lot of discussions between developers on a variety of different topics. These repositories provide developers with perspectives on the issues they face during the apps development process and how they react to them. They play a significant role in improving the capabilities of app developers/end-users and accelerating apps development \cite{mo2015tbil}. Analysing the comments that developers leave in response to the issues might reveal human-centric issues from the viewpoint of developers.


To achieve the goals of this study, we first manually analysed 1,200 app reviews and another 1,200 issue comments collected from 12 mobile app projects that exist on both Google App Store and GitHub. Our analysis led to a taxonomy of human-centric issues. The taxonomy includes three high-level categories: \emph{App Usage}, \emph{Inclusiveness}, and \emph{User Reaction}. The App Usage category consists of \emph{Resource Usage}, \emph{Buginess}, \emph{Change \& Update}, \emph{UI \& UX}, \emph{Privacy \& security}, \emph{Usage instruction}, \emph{Access issues}, and \emph{Others}. The Inclusiveness category covers issues related to \emph{Compatibility}, \emph{Location}, \emph{Language}, \emph{Accessibility}, and \emph{Others}. Finally, the User Reaction category consists of \emph{Fulfilling interests}, \emph{Emotional aspects}, \emph{Preference}, and \emph{Others}. Our analysis found that there are almost twice as many human-centric issues reported in app reviews to issue comments. End-users report more App Usage related issues, followed by User Reaction and Inclusiveness. Developers discuss more App Usage issues, and equally Inclusiveness and User Reaction. We also developed several machine learning (ML) and Deep Learning (DL) models with promising results to automatically detect human-centric issues from app reviews and issue comments. Finally, we conducted an online survey with 16 software/app practitioners, showing the usefulness of automated detection of human-centric issues in app reviews and issue comments in practice.

Some preliminary results of this study were published in \cite{khalajzadeh2022diverse}. In this paper, we build on top of this previous study by extending the first Research Question (\textbf{RQ1}) and adding two new RQs (\textbf{RQ2} and \textbf{RQ3}). More specifically, we built a whole new dataset with new projects compared to the dataset and projects used in \cite{khalajzadeh2022diverse}, leading to a more comprehensive taxonomy of human-centric issues in apps.

The main contributions of this work include: 
\begin{itemize}
\item Developing a taxonomy of human-centric issues discussed in different GitHub repositories and Google App Store reviews;
\item Developing ML and DL based models to detect and classify human-centric issues in app reviews and GitHub issue comments;
\item Providing some implications and possible future research directions to better manage human-centric issues in the software development process;
\item Collecting and analysing developers' viewpoint on the usefulness of using an automated way to detect human-centric issues; and
\item Building and publicly releasing a replication package to enable researchers and practitioners to access all collected data and replicate and validate our study \cite{anonymous_2021_4739069}.
\end{itemize}


The rest of the paper is structured as follows. Section \ref{sec:background/motivation} provides a motivating example for this research. Section \ref{sec:researchmethod} presents our research methodology. Sections \ref{sec:findings} - \ref{sec:RQ3appfindings} present the approaches and results related to our three RQs. Section \ref{sec:Discussion} discusses and reflects on the key findings. In Section \ref{sec:ThreatsValidity}, we list the possible threats to the validity of our study. Finally, Section \ref{sec:RelatedWork} reviews key related work, and Section \ref{sec:Conclusion} draws conclusions and proposes avenues for future work.

\section{Motivation}
\label{sec:background/motivation}

Human aspects -- including \emph{age}, \emph{gender}, \emph{culture}, \emph{language and location}, \emph{digital literacy}, \emph{physical and mental impairments}, and \emph{differing personalities and preferences} -- play an essential role in the uptake of software \cite{grundy2020humanise}. Implications of their negligence are detailed in \cite{khalajzadeh2022diverse}. Consider as a motivating example a dyslexic person who wants to access a website to get some information on their diet. This user has specific requirements to be able to access the website content. As one of the most popular software repositories, GitHub issue tracker provides an option for end-users and developers to report issues and provide feedback on a software system (e.g., a diet website) hosted on GitHub.
A discussion in the issue tracker initiates with a title (\emph{issue title}), followed by subsequent posts (\emph{issue comments}) from reporters and contributors, including project maintainers, developers, users, or the reporter itself. 

\begin{figure}[h!]
\centering
\includegraphics[width=0.61\linewidth]{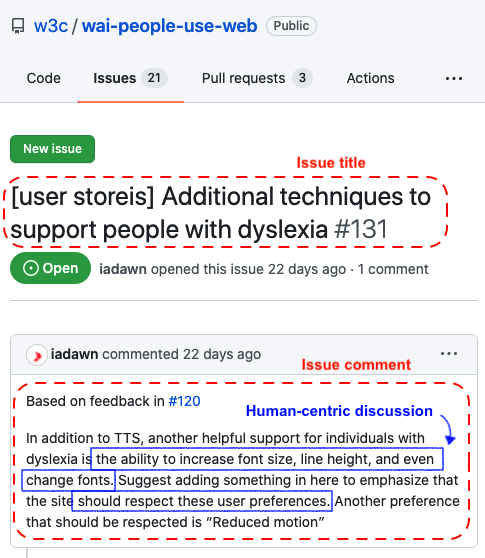}
\vspace{-3mm}
\caption{An example human-centric issue from GitHub}

\label{fig:Motivation}
\end{figure}

Figure \ref{fig:Motivation} shows such an issue in the GitHub issue tracking system made by a collaborator to discuss the dyslexic user's preferences. This is followed by a comment from another collaborator listing some other barriers and asking whether this is true: \textit{``I'm not sure I agree with it''}. The reason for such disagreement is probably that the developer is not fully aware of the needs and preferences of the user. However, discussing such issues can help developers be aware of such challenges and consider such issues when designing software. 
This example shows the importance of paying attention to and discussing issues related to human aspects (i.e., we refer to such issues in this paper as \emph{human-centric issues}) in the uptake of the software. 

\begin{figure}[h!]
\centering
\includegraphics[width=\linewidth]{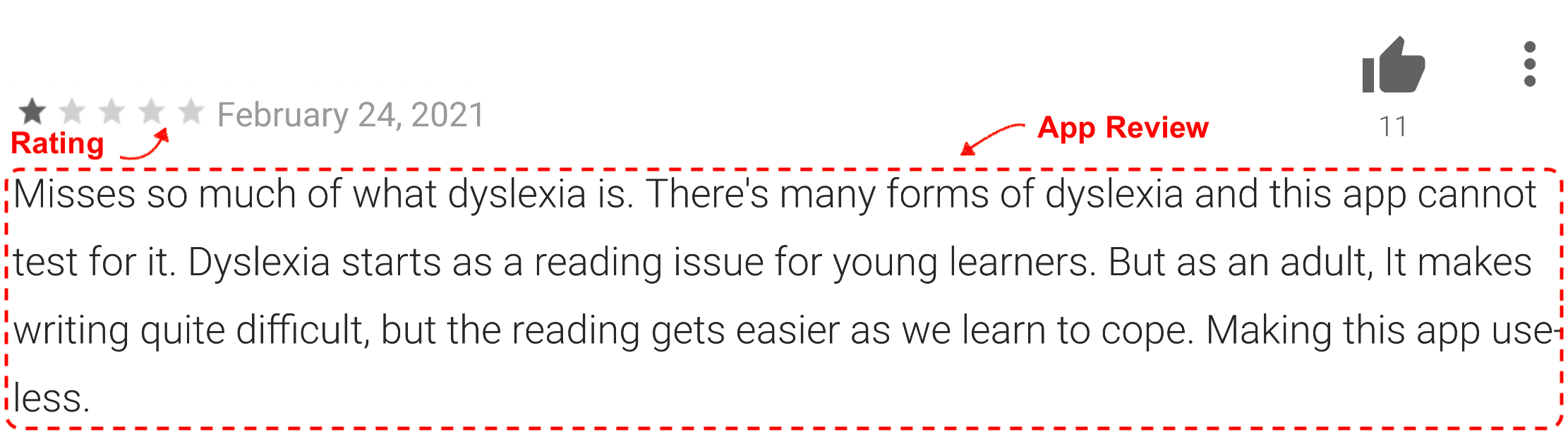}
\vspace{-3mm}
\caption{An example human-centric issue on Google Play Store}

\label{fig:AppReview}
\end{figure}

App reviews, on the other hand, reflect on such aspects from the point of view of the end-users of the system. An app review includes a \emph{comment} and a \emph{rating} posted by the user, and the \emph{number of likes} it has received from other users reading the comment. An example of an app review posted for Dyslexia Reading Test App by a dyslexic user is shown in Figure \ref{fig:AppReview}. Even though the app is designed for a dyslexic user, it is reported as useless by a user with a specific type of dyslexia. This example shows the importance of being aware of the needs of the users, which is not necessarily what a developer assumes. 
Awareness of and discussing human-centric issues may lead to designing more inclusive software for diverse users. 
Such issues are not limited to the users with special needs. As another example, if an app has compatibility issues, it excludes users with a specific device or software from using it. If an app does not provide different languages, it excludes the users who do not understand the provided languages. 
Therefore, there is a need for better understanding, supporting, and promoting awareness of human-centric issues to be able to be able to design more inclusive software. 

\section{Research Method}
\label{sec:researchmethod}

\begin{table*}
\centering
\caption{List of projects studied in this paper. Number of Issues (\faExclamationCircle); Number of Contributors (\faGroup); Number of Stars (\faStarO); Number of Forks (\faCodeFork); Number of Downloads in millions (\faDownload)}
\label{tbl:projectslist}
{\scriptsize
\fontsize{8.5}{9}
\renewcommand{\arraystretch}{1.5}
\begin{tabular}{|l|l|l|c|c|c|c|c|c|}
\hline
\textbf{Project Name}                                    & \textbf{GitHub Repository}                                    & \textbf{Google Play-Store ID}                                & 
\textbf{\faExclamationCircle}          & \textbf{\faGroup}       & \textbf{\faStarO}         & \textbf{\faCodeFork}     & \textbf{\faDownload}    \\ \hline \hline
Signal & WhisperSystems/Signal-Android & org.thoughtcrime.securesms 
& 8,871                &    230            &    20,700	& 4,900   & 50+ \\ \hline
Bitcoin-wallet & bitcoin-wallet/bitcoin-wallet & de.schildbach.wallet 
& 522                &   29             &     2,450	& 1,600    &      5+     \\ \hline
Brave & brave/browser-android & com.brave.browser 
& 819                & 11               &     1,028	& 186    &     10+      \\ \hline
Duckduckgo & duckduckgo/android & com.duckduckgo.mobile.android 
& 354                &  52              &     1,948	& 564    &      10+     \\ \hline
Termux & termux/termux-app  & com.termux 
& 1,743                &   54             &    7,900	& 1,200   &    10+       \\ \hline
Fbreader & geometer/FBReaderJ  & org.geometerplus.zlibrary.ui.android 
& 320                &    38            &    1,719	& 805      &     10+      \\ \hline
K-9 & k9mail/k-9 & com.fsck.k9 
& 3,118                & 226              & 5,800	& 2,200      & 5+         \\ \hline
Pixel-dungeon & watabou/pixel-dungeon & com.watabou.pixeldungeon 
& 66                & 1              & 2,788	& 1,000+      &  5+      \\ \hline
Firefox & mozilla-mobile/fenix & org.mozilla.firefox 
& 14,637                & 231              & 5,417 &	1,000+     & 100+        \\ \hline
WordPress & wordpress-mobile/WordPress-Android & org.wordpress.android 
& 6,468                & 149              & 2,478 &	1,200     & 10+       \\ \hline
Cgeo & cgeo/cgeo & cgeo.geocaching 
& 6,726                & 117           & 1,108 &	521      & 5+       \\ \hline
Osmand & osmandapp/Osmand & net.osmand 
& 7,568                &  746           & 22,674	& 817      & 5+       \\ \hline
\end{tabular}
}
\end{table*}


Our study is motivated by the need to help practitioners and researchers be more aware of different types of end-user human-centric issues impacting software and to help identify possible areas for improvement and investment. We hope this would ultimately help us design software that better meets diverse end-users needs.
To be able to achieve these, we formulated the following RQs:

\begin{center}
\begin{tcolorbox}[colback=white!2!white,colframe=black!75!black]
\textbf{RQ1}. {What end-user human-centric issues typically manifest in mobile apps?}
\end{tcolorbox}
\end{center}
\textbf{Motivation}. Answers to \textbf{RQ1} will provide insights into different types of end-user human-centric issues in mobile apps. Previous research has extensively mined developer discussions in issue tracking systems and app reviews to extract different types of issues (e.g., socio-technical issues) in the software development process and software products. Hence, we use app reviews and developer discussions to identify and classify end-user human-centric issues.

\begin{center}
\begin{tcolorbox}[colback=white!2!white,colframe=black!75!black]
\textbf{RQ2.} Can we accurately and automatically classify end-user human-centric issues from developer discussions and app reviews?
\end{tcolorbox}
\end{center}
\textbf{Motivation}. {Although the answer to RQ1 informs app developers of different types of human-centric issues that may stem from an app, they still need to read the entire app reviews/issue comments of a mobile app to understand which of them include discussions on human-centric issues. If so, they also need to determine which types of human-centric issues are discussed in the app reviews and issue comments. Manually identifying and classifying human-centric issues might be tedious, time-consuming, and error-prone for app developers. An accurate and automated approach can help developers. To answer \textbf{RQ2}, we experiment with several ML/DL methods to automatically and accurately classify such issues in developer discussions and app reviews.} 


\begin{center}
\begin{tcolorbox}[colback=white!2!white,colframe=black!75!black]
\textbf{RQ3.} Do practitioners perceive that the automated classification of end-user human-centric issues in apps is useful?
\end{tcolorbox}
\end{center}
\textbf{Motivation}. {The automated classification of human-centric issues would be useful only if app developers perceive this process useful. We conduct an online survey to gather feedback from developers on the usefulness of the automated classification of human-centric issues. More specifically, as the survey respondents are not going to use the developed automated approaches in RQ2, our survey, similar to other studies \cite{nasab2021automated,prana2019categorizing, abualhaija2020automated}, only seeks developers' opinions about possible benefits that the automated classification of human-centric issues might bring to mobile app development.}

\begin{figure*}[t]
\centering
\includegraphics[scale=0.75]{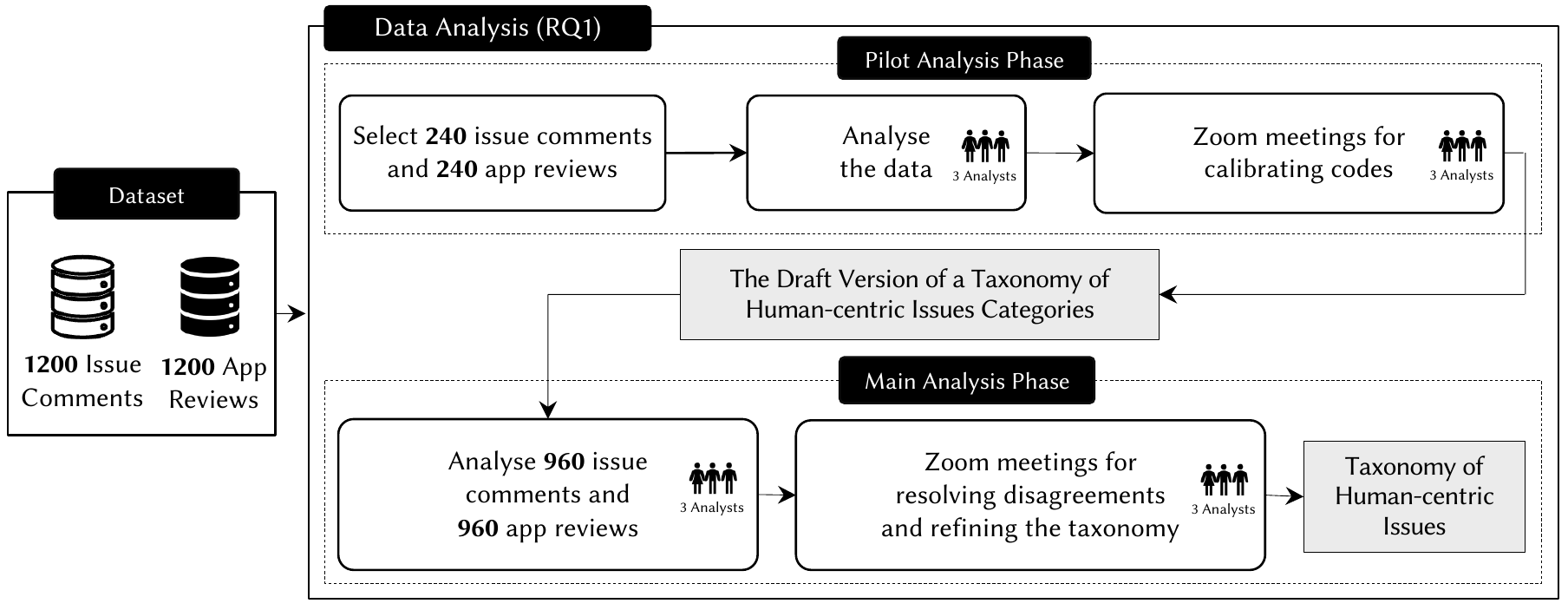}
\vspace{-3mm}
\caption{An overview of data analysis for RQ1} 
\vspace{-0.25cm}
\label{fig:researchmethodRQ1}
\end{figure*}

\subsection{Data Collection}

We needed a subset of open-source mobile apps to answer \textbf{RQ1} and \textbf{RQ2}. The issue tracking systems of such open-source mobile apps had to be available to access their developer discussions.  At the same time, such mobile apps needed to be available on one of the popular mobile app stores to access their user reviews.

We leveraged a dataset of open-source mobile apps created by Mazuera-Rozo et al. \cite{mazuera2020investigating} for this purpose. Mazuera-Rozo et al. built a dataset of 100 Android and 100 iOS mobile apps hosted on GitHub. These 200 Android and iOS mobile apps were randomly collected from an extensive collection of Android and iOS open-source apps identified through the open-source-android-apps project \cite{Androidapplist} and the open-source-ios-apps project \cite{iOSapplist}. Given the qualitative nature of \textbf{RQ1} and the lack of existing datasets of developer discussions and app reviews that include human-centric issues (\textbf{RQ2}), we applied a set of criteria on Mazuera-Rozo et al.'s dataset to sample a representative set of apps. 
\begin{itemize}
    \item \textbf{Android apps}. We only focused on Android apps among the top 100 Android \cite{li2021step} and 100 iOS mobile apps in Mazuera-Rozo et al.'s dataset \cite{mazuera2020investigating}. 
    \item \textbf{Number of issue comments}. We further restricted our selection to Android apps with more than 100 issue comments, including closed and open issues in GitHub.
    \item \textbf{Number of stars.} We decided to opt for Android apps with more than 1,000 stars.
    \item \textbf{Available on the Google Play Store}. Android apps had to have an account on the Google Play Store and were available for Android users to post app reviews.
    \item \textbf{Number of downloads}. Finally, we only selected Android apps with more than 5 million downloads from the Google Play Store.
\end{itemize}

Applying these criteria led to reducing the number of Android apps in Mazuera-Rozo et al.'s dataset from 100 to 12 Android apps. Table \ref{tbl:projectslist} provides an overview of these 12 Android apps. We randomly collected 1,200 developer discussions (issue comments) from GitHub's issue tracking systems of these 12 apps. Each issue can include one or more issue comments. The 1,200 issue comments comprised 10 of 100 randomly selected issue comments from 12 Android apps. Similarly, we randomly selected 100 app reviews from each Android app to generate 1,200 app reviews.

\section{Categories of End-user Human-centric Issues (RQ1)}
\label{sec:findings}

\subsection{Approach}
As shown in Figure \ref{fig:researchmethodRQ1}, the data analysis for RQ1 was conducted in two phases:

\subsubsection{Pilot Analysis Phase} We randomly chose 240 issue comments and 240 app reviews from 2,400 issue comments and app reviews, 20 app reviews and 20 issue comments from each project. The first three authors (analysts) independently conducted the open coding technique \cite{glaser2017discovery} to analyse the data. After finishing the coding process, they held several Zoom meetings to discuss the generated codes, identify duplicates, calibrate the codes, and resolve disagreements. This process resulted in the development of a draft version of a taxonomy of human-centric issues, which grouped human-centric issues into three categories: \emph{App Usage}, \emph{Inclusiveness}, and \emph{User Reaction}. The App Usage category was further divided into \emph{Resource Usage}, \emph{Buginess}, \emph{Change/Update}, \emph{UI/UX}, \emph{Privacy/Security}, \emph{Usage Instruction}, and \emph{Access Issues}. We also classified the Inclusiveness category into four subcategories: \emph{Compatibility}, \emph{Location}, \emph{Language}, and \emph{Accessibility}. Similarly, the User Reaction category was divided into \emph{Fulfilling Interests}, \emph{Emotional Aspects}, and \emph{Preferences} subcategories. {Our pilot analysis indicated that 185 (125 app reviews + 60 issue comments) out of the investigated 480 issue comments/app reviews included at least one human-centric issue. The pilot analysis phase led to three important observations:}

{\textbf{\textit{Observation 1}}. The human-centric categories and subcategories are not mutually exclusive as an issue comment or app review could be coded with more than one human-centric issue category/subcategory. For example, the following app review was coded as “\textit{App Usage}” and “\textit{Inclusiveness}”: “\textit{The lack of universal night mode is the only thing preventing me from keep using this, the blinding white background on some sites are just unbearable.}”}

{\textbf{\textit{Observation 2}}. The labelling process should not be driven by keywords. For example, we found many issue comments/app reviews that had bug-related words, but we did not label them in the “\textit{Buginess}” category and vice versa. As an example, the following review was labelled as “\textit{Buginess}”, but it does not include bug-related keywords. “\textit{It’s nice to be able to see the ones the official geocaching apps make you pay for, but I can’t see any way to log anything. After writing my log entry and selecting the date, I push the only button that looks like submit and it says it’s downloading data, but I need to check my internet connection. I’m definitely online, so I don’t know what’s up.}”}

{\textbf{\textit{Observation 3}}. Human-centric issues categories are not mutually exclusive to the technical categories in the literature, e.g., a review containing a technical bug report may also contain a human-centric issue. We found reviews/comments that purely discussed technical bugs and did not label them as “\textit{Buginess}”, e.g. “\textit{The bug is not resolved and very annoying}” is considered a non-human-centric issue. Taking our definition of human-centric issues as a foundation for identifying human-centric issues in issue comments/app reviews, if a user discussed a technical issue in issue comments/app reviews that stems from the lack of (proper) consideration of their specific/differing characteristics, limitations, abilities, and personalities, we considered that comment/review as a human-centric issue. For example, the “\textit{Buginess}” category should refer to any issues discussing the bugs an app has, which impacts the usage of the app by different users. The same process should be applied to the rest of the categories of human-centric issues in the taxonomy.}

\subsubsection{Main Analysis Phase}
In this phase, the same analysts in the pilot phase analysed the rest of 1,920 issue comments/app reviews in several iterations. Each of the analysts manually analysed 640 individual issue comments/app reviews. In other words, each issue comment was analysed and labeled by two analysts. In each iteration, the analysts analysed approximately 200 issue comments or app reviews. We created a spreadsheet and shared it with three analysts based on the taxonomy created in the pilot analysis phase. The analysts were asked to indicate whether an issue comment or an app review included one or more human-centric issues. If so, they had to specify which of the categories/subcategories in the taxonomy of human-centric issues the given issue comment or app review belonged to and put ``1'' in the corresponding columns in the spreadsheet. The analysts had the freedom to capture and add any new human-centric issues category/subcategory to the taxonomy if they were not captured in the draft version of the taxonomy. 


\begin{figure*}[t]
\centering
\includegraphics[width=0.9\linewidth]{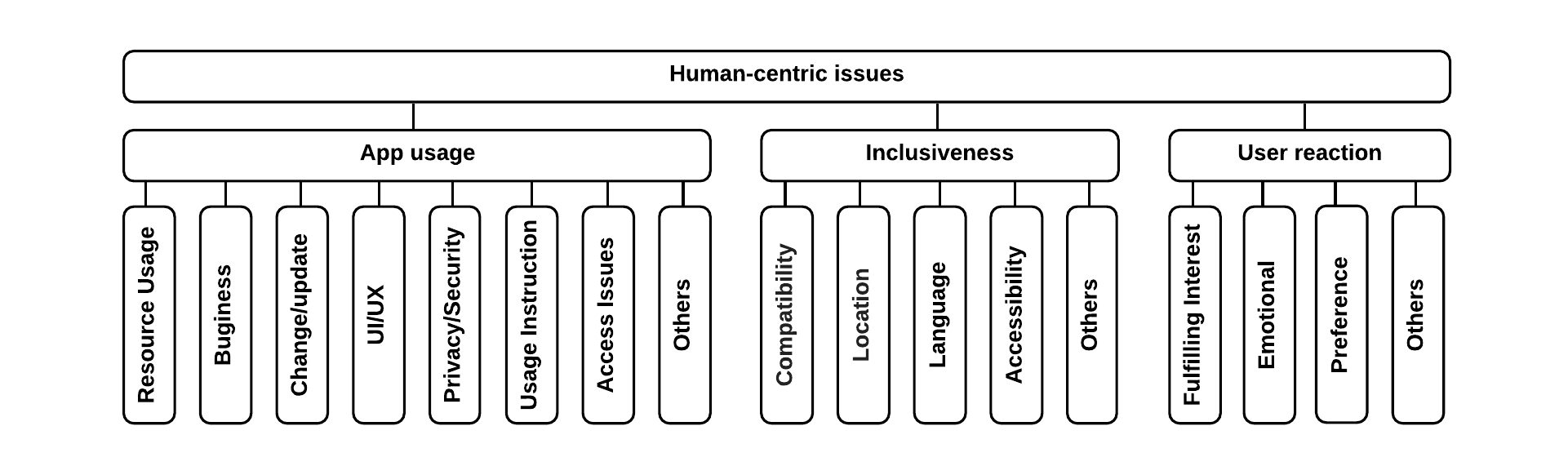}
\vspace{-3mm}
\caption{A taxonomy of end-user human-centric issues}
\label{fig:taxonomy}
\end{figure*}

At the end of each iteration, the three analysts had a roughly 1-hour Zoom meeting to compare their labelling results and resolve possible disagreements. The majority of disagreements were resolved through discussions between the two assigned analysts by explicitly providing the reason behind their choices. The third analyst was asked to read and label the conflicting issue comment if the two analysts could not agree. Then, we voted to resolve the disagreement. {We had a total of 49 issue comments among 1,200 issue comments and 56 app reviews among 1,200 app reviews that we further discussed. The reason for discussing them was not always that we did not come up with the same categories, sometimes because one of the analysers flagged the issue comment/app review for further clarification and discussion. We were able to reach an agreement for all of them after the initial discussion. 
}

We found a few new human-centric issues compared to the draft version of the taxonomy and grouped them as ``others'' in each category. {We found 3 issue comments related to the “app usage” category that did not belong to any of the sub-categories, 2 in the “inclusiveness” and 1 in the “user reaction”. We categorised them in new categories as the “others” sub-categories under each category. In app reviews, we found 19 app reviews related to the “app usage” category, 3 in the “inclusiveness” and 4 in the “user reaction” that did not belong to any of the sub-categories, and similarly, categorised them in new “others” sub-categories under each category.} The final taxonomy is shown in Figure \ref{fig:taxonomy}. We acknowledge that our process in the main phase does not follow the idea of open coding. We decided this since it could avoid developing a potentially very large number of possible human-centric issue categories \cite{humbatova2020taxonomy}. Also, it supported the analysts to reach and use consistent labelling without introducing substantial bias \cite{humbatova2020taxonomy}. 

\subsection{Findings}
This section presents the taxonomy of human-centric issues. 


\subsubsection{\textbf{App Usage}}
The App Usage category covers all the issues related to the app itself and how it impacts users experiences. 
This category consists of {Resource Usage},	{Buginess}, {Change \& Update}, {UI \& UX}, {Privacy \& Security}, {Usage Instruction}, {Access Issues}, and {Others}. 

\textit{\textbf{Resource Usage}}:
Resource usage refers to the issues end-user have with battery usage, internet connection, server, memory and resource management. An example of Resource Usage issue discussed in GitHub is: 
\begin{quoting}
\noindent\faComment \hspace{0cm} \textit{\textbf{GitHub Issue} }-
\textit{``... continuously tries to detect the postal address...
But in any case, \textbf{the battery drain sounds just too excessive}...}" - (Osmand)

\end{quoting}

A user reports a review of Bitcoin wallet app on Google App Store 
mentioning that ``\textit{there REALLY needs to be a way to \textbf{limit bandwidth} used when syncing with the blockchain.}" The user has bought a new phone, and transferred the wallet backup from the old phone and started the restore process, and is  complaining that ``\textit{It's used \textbf{100GB of data so far in the last 24 hours}, and there are still three YEARS worth of blockchain to go.}" - (Bitcoin wallet)


\textit{\textbf{Buginess}}:
Buginess category refers to any issues discussing the bugs an app has, which impacts the usage of the app by different users. Examples of Buginess issue reported in GitHub and Google App Store are shown below:

\begin{quoting}
\noindent\faComment \hspace{0cm} \textit{\textbf{GitHub Issue} }-
\textit{``\textbf{I had one crash} after long time of using the live map, but was \textbf{unable to debug/reproduce}. Most probably a \textbf{problem} we have had already in the past."}
 - (Cgeo)

\end{quoting}

\begin{quoting}
\noindent\faCommentsO \hspace{0cm} \textit{\textbf{App Review} }-
\textit{``... I find the support to be lacking and the \textbf{bugs to be plentiful}. Good luck getting rid of it though, \textbf{tried closing my account due to number of bugs} and no one could contact me anymore via SMS."} 
- (Signal)

\end{quoting}

\textit{\textbf{Change \& Update}}:
This category refers to any issues that did not exist before, however, users face when updating the app or upgrading to a new version. Users sometimes feel unhappy with the new changes made and prefer or were more comfortable with the older options, or features. An issue discussed on GitHub relates to a user having always
had some options turned off (unchecked), but with the new upgrade, ``\textit{\textbf{these checkbox preferences} don't seem to be being respected when composing a message.}" 
 - (K - 9). Another user complains about a ``\textit{Disastrous update}" through Brave App review, asking how they can
get the \textbf{previous version} since ``\textit{Switch between tabs has become extremely difficult.}"
- (Brave)


\textit{\textbf{UI \& UX}}:
Any discussions related to the User Interface (UI) and User Experiences (UX) with the app are labeled in this category. Examples of UI \& UX issue reported in GitHub and Google App Store are shown below:

\begin{quoting}
\noindent\faComment \hspace{0cm} \textit{\textbf{GitHub Issue} }-
\textit{``We also need to consider the position of the toolbar for custom tabs. The browser is meant to \textbf{look more} like the native app itself to give the illusion of not leaving the context of the calling app, so we might want to think about re-positioning the toolbar to the top like most apps would have."}
 - (Firefox)

\end{quoting}

\begin{quoting}
\noindent\faCommentsO \hspace{0cm} \textit{\textbf{App Review} }-
\textit{``...However please provide a navigation tool to understand the features. Tough for \textbf{not so technical people} to navigate. \textbf{Also please improve your notification icon. Make it catchy.} Most of new users don't even know that there is a message."}
- (Signal)

\end{quoting}

\textit{\textbf{Privacy \& Security}}: This covers issues related to the users' privacy, security, data protection, reliability, and trust concerns. We classified discussions on accessing the location and private data of a user into this category. As an example, a developer has claimed on GitHub, that
``\textit{We frequently audit our browser to \textbf{detect any possible privacy implications}. If you have \textbf{detected any possible security issue}, reach us via hackerone, it will go through our privacy expert team...}" - (Duckduckgo). 
However, a user reported that ``\textit{Duckduckgo
is \textbf{advertised as a security and privacy web browser}, yet there are almost no advanced options to customise those aspects. What about https, dns, and script options? or control over what information is shared back to you guys?}" - (Duckduckgo)


\textit{\textbf{Usage Instruction}}:
Some apps do not provide enough instructions for the users to be able to use it. Lack of usage instruction leads to some confusion for the users. We report all such issues in this category. For example, a developer discusses how to resolve such issue on GitHub:

\begin{quoting}
\noindent\faComment \hspace{0cm} \textit{\textbf{GitHub Issue} }-
\textit{``I am going to extrapolate here and guess that the average user will find the same problem, and may never find this context menu. \textbf{To solve that, we could either introduce a UI hint on first use}, or make it more hangouts-y..."}
 - (Signal)

\end{quoting}

And a user reports their experience using an app lacking enough documentation and instructions:

\begin{quoting}
\noindent\faCommentsO \hspace{0cm} \textit{\textbf{App Review} }-
\textit{``There are several map rendering styles but unfortunately I was not able to find one that rendered the map in a way I liked for use when walking in GB. It is possible to create ones own rendering style which I have done \textbf{but documentation on how to do so is appalling and support from OsmAnd was lacklustre.}"}
- (Osmand)

\end{quoting}

\textit{\textbf{Access Issues}}: Users sometimes have issues with downloading, registering and accessing and app. All these issues are categorised in Access issues category. There were a lot of app reviews reporting Access issues, such as a user reporting ``\textit{The app does not install in my phone please any advice to install in phone.}"
- (Termux). However, we did not find any GitHub issues discussing how to deal with such an issue.





\textit{\textbf{Others}}: Finally, we categorise all other issues related to the usage of the app, that are not directly related to the previous categories, in the Others category. Examples of Others issue reported in GitHub and Google App Store are shown below. The first example discusses how the developer thinks users expect/prefer and how to react accordingly, and the second one shows the lack of trust by the user to the app.

\begin{quoting}
\noindent\faComment \hspace{0cm} \textit{\textbf{GitHub Issue} }-
\textit{ ``I believe that we should sync the deletion of individual pages but that we shouldn't sync bulk deletion like "Clear All".
Given mobile devices will have synced desktop History, a "Clear All" will wipe your desktop History on your local mobile device but we don't think \textbf{users would expect it to also wipe all of the History on their desktops too.}"}
 - (Firefox - Android)

\end{quoting}

\begin{quoting}
\noindent\faCommentsO \hspace{0cm} \textit{\textbf{App Review} }-
\textit{``...This is my 3rd month claim, brave is \textbf{fooling there users}. Or they \textbf{never added support programme for users}.Don't install."}
- (Brave - Android)

\end{quoting}

\subsubsection{\textbf{Inclusiveness}}
This category covers the issues related to the inclusion, exclusion or discrimination toward a specific groups of users. It includes issues related to the age, gender, and socio-economic status of the users. We categorise Inclusiveness into five different subcategories, as: {Compatibility}, {Location}, {Language}, {Accessibility}, and {Others}.

\textit{\textbf{Compatibility}}:
Any discussions around the compatibility of an app with different devices, operating systems, and platforms are included in this category. Compatibility issues are normally thought of as technical, not human-centric issues. However, a common reason for them occurring can be because of the users' socio-economic status, i.e., not having access to the latest phones, or the developers' ignorance, i.e., not taking all different platform choices into account. Examples of Compatibility related issues reported on GitHub and Google App Store are shown below:

\begin{quoting}
\noindent\faComment \hspace{0cm} \textit{\textbf{GitHub Issue} }-
\textit{``...it appears that \textbf{FbReader 1.6.4 is marked as incompatible on my CM7.2} (Android 2.3.7) phone, even though the maxsdk is 10. FBReader 1.6.1 is compatible. Are we building a wrong branch perhaps?"}
 - (FbReader)

\end{quoting}
A user left a comment through Bitcoin wallet app review, complaining that they want to transfer to another app: 
``\textit{Shame because it's been simple and reliable and easy to backup/restore... \textbf{Just isn't compatible with latest wallet standards.}}"
- (Bitcoin wallet)


\textit{\textbf{Location}}:
This covers issues related to the physical location from where the user is accessing the software. Based on our analysis, users' access may be limited if they are visiting a country and have no local phone number or App store account. A developer discusses on GitHub that 
\textit{``need to be sure the field notes contain the correct z-time...It will only work if \textbf{user sets their time zone }on gc.com as well."}
 - (Cgeo). 
Through app reviews, a user has reported that 
\textit{``The search engine is pretty bad, anything I search for \textbf{only shows results in the USA and beyond}.. Good but I have to put UK at the end of every search."}
- (Duckduckgo)


\textit{\textbf{Language}}:
This category includes discussions about language or culture-related issues. For example, a user asks ``\textit{how about adding in a \textbf{Language Translation program/option} into DuckDuckgo, so you can translate other websites and mostly all \textbf{foreign languages}?}"
- (Duckduckgo). On the other hand, on Duckduckgo GitHub, in response to another language-related issue, a developer discusses that {``\textit{...we don't currently support \textbf{Brazilian Portuguese}.
We would love to \textbf{support as many languages as possible}, but at the same time, we need to keep a manageable size of translations...}"}
 - (Duckduckgo)

\textit{\textbf{Accessibility}}:
This category covers issue comments and app reviews discussing accessibility issues. Discussions about the users with physical and mental impairments also fall into this category. Examples include:

\begin{quoting}
\noindent\faComment \hspace{0cm} \textit{\textbf{GitHub Issue} }-
\textit{``\textbf{If we disable accessibility services, how are disabled users} going to enter their passwords and passphrases?"}
 - (Bitcoin wallet)

\end{quoting}

\begin{quoting}
\noindent\faCommentsO \hspace{0cm} \textit{\textbf{App Review} }-
\textit{``Not intuitive. \textbf{Difficult to change fonts, text size, etc.} Couldn't not figure out how to purchase a book." }
- (FbReader)

\end{quoting}

\textit{\textbf{Others}}:
All other issues related to the inclusiveness of diverse users that are not directly related to the previous categories are considered in the Others category. Examples of Others issue reported in GitHub and Google App Store are shown below. These two examples are related to the technical knowledge and background of the users.

\begin{quoting}
\noindent\faComment \hspace{0cm} \textit{\textbf{GitHub Issue} }-
\textit{``The  reason for this to ship in Firefox itself, instead of an add-on is \textbf{that many users do not know about add-on support.
Most does not even know what an add-on is.} They just install the browser and get on with it.
This should be in Firefox itself, if you ask me!"}
 - (Firefox)

\end{quoting}

\begin{quoting}
\noindent\faCommentsO \hspace{0cm} \textit{\textbf{App Review} }-
\textit{`` Maybe have the general search in the center of the app so people don't think they have to know the exact website address they are looking for in order to use the app.\textbf{ That might be confusing to some who aren't so tech savvy.}" }
- (Duckduckgo)

\end{quoting}

\subsubsection{\textbf{User Reaction}}
This category covers the issues users have when accessing the application due to their specific preferences, interests, and emotional reactions. We further divided this into \textbf{Fulfilling interests}, \textbf{Emotional aspects}, \textbf{Preference}, and \textbf{Others}.

\textit{\textbf{Fulfilling interests}}:
Users use apps for different purposes. Some applications might be suitable for a user based on their goals and objectives, but not for another user with different needs. We categorise discussions related to user satisfaction due to the fulfilment of their interests in this category. For example, an issue comment discusses that due to the use of third-party libraries, a user is 
about to switch eBook viewers. ``\textit{FBReader is currently just not usable on my device, and eBook viewer is the main reason I use the device.}"
 - (FbReader). A user also reports that  
\textit{``Termux isn't a good app to use its a app exec controlled app \textbf{they don't let you use sudo and the commands don't execute as you type them. Don't waste your time on this app."}}
- (Termux). These two examples show how not fulfilling users' interest might disappoint them from using the app.


\textit{\textbf{Emotional Aspects}}:
This category includes the possible emotional impacts that the software can have on the users, including making the users confused, worried, scared and bored. Examples of Emotional Aspects issue reported in GitHub and Google App Store are shown below:

\begin{quoting}
\noindent\faComment \hspace{0cm} \textit{\textbf{GitHub Issue} }-
\textit{``The amount of off-topic content here is \textbf{shameful, and very stressful} for anyone that then has to clean up your mess. I really don't see how you expect a collaborative response when this is the attitude here."}
 - (Duckduckgo)

\end{quoting}

\begin{quoting}
\noindent\faCommentsO \hspace{0cm} \textit{\textbf{App Review} }-
\textit{``After some update the app crashes on some specific access to the contacts \textbf{which is really really really annoying }when you have written a long mail and you loose all of it even in the draft. If the problem isn't fixed I am changing app forever"}
- (K - 9)

\end{quoting}

\textit{\textbf{Preferences}}:
Any discussion related to the user's preferences fall into this category. This relates to the features or functionalities that users prefer based on their specific human characteristics. Preference-related discussions include different aspects: (1) requesting new features (2) issues or requests to change an existing feature, such as the position of user interface elements (3) and privacy-related issues due to personal reasons. Preferences are sometimes discussed according to users' feedback received through app reviews or by developers from the users' perspective.
In some apps, developers often use the app themselves and discuss their usage experiences on GitHub. An examples of Preferences issue discussed in GitHub is 
that the developer has
made some changes in dictionary code 
``\textit{because not all users want to use English - German dictionaries. E.g., \textbf{I prefer English - Russian \& German - Russian}. ;) I think that is good idea to list only \textbf{'universal dictionaries'.}}"
 - (FbReader)

A user on Google App Store complains about Firefox:
``\textit{no option to even download features that make the pc firefox so effective and easy to use. No clear recent search history and Bookmark browsing that takes painfully slow to scroll through, no book mark search feature. \textbf{I would rate this 5 stars but I would prefer Firefox not Firefox lite}}"
- (Firefox)


\textit{\textbf{Others}}:
All other issues related to the reaction of different users based on their diverse characteristics not directly related to the previous categories are considered in the Others category. Bellow shows two examples in the Others category reported in GitHub and Google App Store:

\begin{quoting}
\noindent\faComment \hspace{0cm} \textit{\textbf{GitHub Issue} }-
\textit{``...I will have a look at tomorrows NB and try again from the perspective of \textbf{a first time user}"}
 - (Cgeo)

\end{quoting}

\begin{quoting}
\noindent\faCommentsO \hspace{0cm} \textit{\textbf{App Review} }-
\textit{``\textbf{I can't report an issue} In the app either because every time I press the option to report it just brings up the typical share menu for Android devices where it recommends apps or people to share with."}
- (Bitcoin wallet)

\end{quoting}

\begin{table*}[]
\centering
\caption{Number (\#) and percentage (\%) of human-centric issues in issue comments in our 12 projects}
\label{tbl:TableIssue}
{\scriptsize
\renewcommand{\arraystretch}{1.5}
\begin{tabular}{l|cccccccc|ccccc|cccc|l}
\cline{2-19}
 &
  \multicolumn{8}{c|}{{\color[HTML]{222222} \textbf{\#1: App Usage}}} &
  \multicolumn{5}{c|}{{\color[HTML]{222222} \textbf{\#2: Inclusiveness}}} &
  \multicolumn{4}{c|}{{\color[HTML]{222222} \textbf{\#3: User Reaction}}} &
  \multicolumn{1}{c|}{\multirow{2}{*}{\textbf{\rotatebox[origin=c]{90}{total (Out of 100) }}}} \\ \cline{2-18}
 &
  \multicolumn{1}{l|}{\textbf{\rotatebox[origin=c]{90}{Resource Usage}}} &
  \multicolumn{1}{l|}{\textbf{\rotatebox[origin=c]{90}{Buginess}}} &
  \multicolumn{1}{l|}{\textbf{\rotatebox[origin=c]{90}{Change/Update}}} &
  \multicolumn{1}{l|}{\textbf{\rotatebox[origin=c]{90}{UI/UX}}} &
  \multicolumn{1}{l|}{{\color[HTML]{222222} \textbf{\rotatebox[origin=c]{90}{Privacy/security}}}} &
  \multicolumn{1}{l|}{\textbf{\rotatebox[origin=c]{90}{ Usage instruction }}} &
  \multicolumn{1}{l|}{\textbf{\rotatebox[origin=c]{90}{Access issues}}} &
  \multicolumn{1}{l|}{\textbf{\rotatebox[origin=c]{90}{Others}}} &
  \multicolumn{1}{l|}{{\color[HTML]{222222} \textbf{\rotatebox[origin=c]{90}{Compatibility}}}} &
  \multicolumn{1}{l|}{{\color[HTML]{222222} \textbf{\rotatebox[origin=c]{90}{Location}}}} &
  \multicolumn{1}{l|}{{\color[HTML]{222222} \textbf{\rotatebox[origin=c]{90}{Language}}}} &
  \multicolumn{1}{l|}{\textbf{\rotatebox[origin=c]{90}{Accessibility}}} &
  \multicolumn{1}{l|}{\textbf{\rotatebox[origin=c]{90}{Others}}} &
  \multicolumn{1}{l|}{\textbf{\rotatebox[origin=c]{90}{ Fulfilling interests }}} &
  \multicolumn{1}{l|}{{\color[HTML]{222222} \textbf{\rotatebox[origin=c]{90}{ Emotional aspects }}}} &
  \multicolumn{1}{l|}{{\color[HTML]{222222} \textbf{\rotatebox[origin=c]{90}{Preference}}}} &
  \multicolumn{1}{l|}{{\color[HTML]{222222} \textbf{\rotatebox[origin=c]{90}{Others}}}} &
  \multicolumn{1}{l|}{} \\ \cline{19-19} \hline
\multicolumn{1}{|l|}{\textbf{Signal}} &
  \multicolumn{1}{c|}{5} &
  \multicolumn{1}{c|}{14} &
  \multicolumn{1}{c|}{2} &
  \multicolumn{1}{c|}{2} &
  \multicolumn{1}{c|}{5} &
  \multicolumn{1}{c|}{2} &
  \multicolumn{1}{c|}{0} &
  1 &
  \multicolumn{1}{c|}{3} &
  \multicolumn{1}{c|}{0} &
  \multicolumn{1}{c|}{1} &
  \multicolumn{1}{c|}{5} &
  0 &
  \multicolumn{1}{c|}{0} &
  \multicolumn{1}{c|}{4} &
  \multicolumn{1}{c|}{4} &
  0 & \multicolumn{1}{c|}{35} \\ \hline
\multicolumn{1}{|l|}{\textbf{Cgeo}} &
  \multicolumn{1}{c|}{2} &
  \multicolumn{1}{c|}{7} &
  \multicolumn{1}{c|}{3} &
  \multicolumn{1}{c|}{7} &
  \multicolumn{1}{c|}{0} &
  \multicolumn{1}{c|}{3} &
  \multicolumn{1}{c|}{0} &
  0 &
  \multicolumn{1}{c|}{5} &
  \multicolumn{1}{c|}{1} &
  \multicolumn{1}{c|}{2} &
  \multicolumn{1}{c|}{2} &
  0 &
  \multicolumn{1}{c|}{2} &
  \multicolumn{1}{c|}{0} &
  \multicolumn{1}{c|}{8} &
  1 & \multicolumn{1}{c|}{33} \\ \hline
\multicolumn{1}{|l|}{\textbf{Firefox}} &
  \multicolumn{1}{c|}{3} &
  \multicolumn{1}{c|}{1} &
  \multicolumn{1}{c|}{0} &
  \multicolumn{1}{c|}{9} &
  \multicolumn{1}{c|}{7} &
  \multicolumn{1}{c|}{0} &
  \multicolumn{1}{c|}{0} &
  1 &
  \multicolumn{1}{c|}{1} &
  \multicolumn{1}{c|}{0} &
  \multicolumn{1}{c|}{1} &
  \multicolumn{1}{c|}{9} &
  1 &
  \multicolumn{1}{c|}{0} &
  \multicolumn{1}{c|}{1} &
  \multicolumn{1}{c|}{10} &
  0 & \multicolumn{1}{c|}{30} \\ \hline
\multicolumn{1}{|l|}{\textbf{Duckduckgo}} &
  \multicolumn{1}{c|}{2} &
  \multicolumn{1}{c|}{8} &
  \multicolumn{1}{c|}{1} &
  \multicolumn{1}{c|}{6} &
  \multicolumn{1}{c|}{8} &
  \multicolumn{1}{c|}{1} &
  \multicolumn{1}{c|}{0} &
  1 &
  \multicolumn{1}{c|}{0} &
  \multicolumn{1}{c|}{0} &
  \multicolumn{1}{c|}{3} &
  \multicolumn{1}{c|}{0} &
  0 &
  \multicolumn{1}{c|}{1} &
  \multicolumn{1}{c|}{1} &
  \multicolumn{1}{c|}{6} &
  0 & \multicolumn{1}{c|}{29} \\ \hline
\multicolumn{1}{|l|}{\textbf{K-9}} &
  \multicolumn{1}{c|}{2} &
  \multicolumn{1}{c|}{8} &
  \multicolumn{1}{c|}{3} &
  \multicolumn{1}{c|}{6} &
  \multicolumn{1}{c|}{4} &
  \multicolumn{1}{c|}{0} &
  \multicolumn{1}{c|}{0} &
  0 &
  \multicolumn{1}{c|}{4} &
  \multicolumn{1}{c|}{0} &
  \multicolumn{1}{c|}{1} &
  \multicolumn{1}{c|}{0} &
  0 &
  \multicolumn{1}{c|}{2} &
  \multicolumn{1}{c|}{0} &
  \multicolumn{1}{c|}{6} &
  0 & \multicolumn{1}{c|}{29} \\ \hline
\multicolumn{1}{|l|}{\textbf{WordPress}} &
  \multicolumn{1}{c|}{0} &
  \multicolumn{1}{c|}{6} &
  \multicolumn{1}{c|}{0} &
  \multicolumn{1}{c|}{12} &
  \multicolumn{1}{c|}{0} &
  \multicolumn{1}{c|}{0} &
  \multicolumn{1}{c|}{0} &
  0 &
  \multicolumn{1}{c|}{1} &
  \multicolumn{1}{c|}{0} &
  \multicolumn{1}{c|}{2} &
  \multicolumn{1}{c|}{0} &
  0 &
  \multicolumn{1}{c|}{0} &
  \multicolumn{1}{c|}{0} &
  \multicolumn{1}{c|}{6} &
  0 & \multicolumn{1}{c|}{26} \\ \hline
\multicolumn{1}{|l|}{\textbf{Brave}} &
  \multicolumn{1}{c|}{1} &
  \multicolumn{1}{c|}{12} &
  \multicolumn{1}{c|}{0} &
  \multicolumn{1}{c|}{12} &
  \multicolumn{1}{c|}{0} &
  \multicolumn{1}{c|}{1} &
  \multicolumn{1}{c|}{0} &
  0 &
  \multicolumn{1}{c|}{2} &
  \multicolumn{1}{c|}{0} &
  \multicolumn{1}{c|}{0} &
  \multicolumn{1}{c|}{0} &
  0 &
  \multicolumn{1}{c|}{0} &
  \multicolumn{1}{c|}{0} &
  \multicolumn{1}{c|}{4} &
  0 & \multicolumn{1}{c|}{24} \\ \hline
\multicolumn{1}{|l|}{\textbf{Osmand}} &
  \multicolumn{1}{c|}{3} &
  \multicolumn{1}{c|}{2} &
  \multicolumn{1}{c|}{0} &
  \multicolumn{1}{c|}{6} &
  \multicolumn{1}{c|}{1} &
  \multicolumn{1}{c|}{6} &
  \multicolumn{1}{c|}{0} &
  0 &
  \multicolumn{1}{c|}{2} &
  \multicolumn{1}{c|}{0} &
  \multicolumn{1}{c|}{3} &
  \multicolumn{1}{c|}{0} &
  0 &
  \multicolumn{1}{c|}{1} &
  \multicolumn{1}{c|}{0} &
  \multicolumn{1}{c|}{10} &
  0 & \multicolumn{1}{c|}{23} \\ \hline
\multicolumn{1}{|l|}{\textbf{Termux}} &
  \multicolumn{1}{c|}{2} &
  \multicolumn{1}{c|}{5} &
  \multicolumn{1}{c|}{1} &
  \multicolumn{1}{c|}{6} &
  \multicolumn{1}{c|}{5} &
  \multicolumn{1}{c|}{1} &
  \multicolumn{1}{c|}{0} &
  0 &
  \multicolumn{1}{c|}{1} &
  \multicolumn{1}{c|}{0} &
  \multicolumn{1}{c|}{0} &
  \multicolumn{1}{c|}{1} &
  0 &
  \multicolumn{1}{c|}{0} &
  \multicolumn{1}{c|}{0} &
  \multicolumn{1}{c|}{1} &
  0 & \multicolumn{1}{c|}{22} \\ \hline
\multicolumn{1}{|l|}{\textbf{Bitcoin}} &
  \multicolumn{1}{c|}{2} &
  \multicolumn{1}{c|}{6} &
  \multicolumn{1}{c|}{1} &
  \multicolumn{1}{c|}{2} &
  \multicolumn{1}{c|}{6} &
  \multicolumn{1}{c|}{1} &
  \multicolumn{1}{c|}{0} &
  0 &
  \multicolumn{1}{c|}{0} &
  \multicolumn{1}{c|}{0} &
  \multicolumn{1}{c|}{1} &
  \multicolumn{1}{c|}{3} &
  1 &
  \multicolumn{1}{c|}{0} &
  \multicolumn{1}{c|}{0} &
  \multicolumn{1}{c|}{2} &
  0 & \multicolumn{1}{c|}{19} \\ \hline
\multicolumn{1}{|l|}{\textbf{Fbreader}} &
  \multicolumn{1}{c|}{0} &
  \multicolumn{1}{c|}{2} &
  \multicolumn{1}{c|}{0} &
  \multicolumn{1}{c|}{4} &
  \multicolumn{1}{c|}{1} &
  \multicolumn{1}{c|}{0} &
  \multicolumn{1}{c|}{0} &
  0 &
  \multicolumn{1}{c|}{1} &
  \multicolumn{1}{c|}{0} &
  \multicolumn{1}{c|}{7} &
  \multicolumn{1}{c|}{0} &
  0 &
  \multicolumn{1}{c|}{2} &
  \multicolumn{1}{c|}{0} &
  \multicolumn{1}{c|}{5} &
  0 & \multicolumn{1}{c|}{18} \\ \hline
\multicolumn{1}{|l|}{\textbf{Pixel-dungeon}} &
  \multicolumn{1}{c|}{0} &
  \multicolumn{1}{c|}{0} &
  \multicolumn{1}{c|}{0} &
  \multicolumn{1}{c|}{1} &
  \multicolumn{1}{c|}{0} &
  \multicolumn{1}{c|}{0} &
  \multicolumn{1}{c|}{0} &
  0 &
  \multicolumn{1}{c|}{3} &
  \multicolumn{1}{c|}{0} &
  \multicolumn{1}{c|}{11} &
  \multicolumn{1}{c|}{0} &
  0 &
  \multicolumn{1}{c|}{0} &
  \multicolumn{1}{c|}{0} &
  \multicolumn{1}{c|}{1} &
  0 & \multicolumn{1}{c|}{16} \\ \hline
\multicolumn{1}{|l|}{\textbf{Total (\#)}} &
  \multicolumn{1}{c|}{22} &
  \multicolumn{1}{c|}{71} &
  \multicolumn{1}{c|}{11} &
  \multicolumn{1}{c|}{73} &
  \multicolumn{1}{c|}{37} &
  \multicolumn{1}{c|}{15} &
  \multicolumn{1}{c|}{0} &
  3 &
  \multicolumn{1}{c|}{23} &
  \multicolumn{1}{c|}{1} &
  \multicolumn{1}{c|}{32} &
  \multicolumn{1}{c|}{20} &
  2 &
  \multicolumn{1}{c|}{8} &
  \multicolumn{1}{c|}{6} &
  \multicolumn{1}{c|}{63} &
  1 & \\ \cline{1-18} 
\multicolumn{1}{|l|}{\textbf{Total (\%)}} &
  \multicolumn{1}{c|}{1.83} &
  \multicolumn{1}{c|}{5.92} &
  \multicolumn{1}{c|}{0.92} &
  \multicolumn{1}{c|}{6.08} &
  \multicolumn{1}{c|}{3.08} &
  \multicolumn{1}{c|}{1.25} &
  \multicolumn{1}{c|}{0.00} &
  0.25 &
  \multicolumn{1}{c|}{1.92} &
  \multicolumn{1}{c|}{0.08} &
  \multicolumn{1}{c|}{2.67} &
  \multicolumn{1}{c|}{1.67} &
  0.17 &
  \multicolumn{1}{c|}{0.67} &
  \multicolumn{1}{c|}{0.50} &
  \multicolumn{1}{c|}{5.25} &
  0.08 & \\ \cline{1-18} 
\end{tabular}
}
\end{table*}
\begin{table*}[]
\centering
\caption{Number (\#) and percentage (\%) of human-centric issues in app reviews in our 12 projects}
\label{tbl:TableApp}
{\scriptsize
\renewcommand{\arraystretch}{1.5}
\begin{tabular}{l|cccccccc|ccccc|cccc|l}
\cline{2-19}
 &
  \multicolumn{8}{c|}{{\color[HTML]{222222} \textbf{\#1: App Usage}}} &
  \multicolumn{5}{c|}{{\color[HTML]{222222} \textbf{\#2: Inclusiveness}}} &
  \multicolumn{4}{c|}{{\color[HTML]{222222} \textbf{\#3: User Reaction}}} &
  \multicolumn{1}{c|}{\multirow{2}{*}{\textbf{\rotatebox[origin=c]{90}{total (Out of 100)  }}}} \\ \cline{2-18}
 &
  \multicolumn{1}{l|}{\textbf{\rotatebox[origin=c]{90}{Resource Usage}}} &
  \multicolumn{1}{l|}{\textbf{\rotatebox[origin=c]{90}{Buginess}}} &
  \multicolumn{1}{l|}{\textbf{\rotatebox[origin=c]{90}{Change/Update}}} &
  \multicolumn{1}{l|}{\textbf{\rotatebox[origin=c]{90}{UI/UX}}} &
  \multicolumn{1}{l|}{{\color[HTML]{222222} \textbf{\rotatebox[origin=c]{90}{Privacy/security}}}} &
  \multicolumn{1}{l|}{\textbf{\rotatebox[origin=c]{90}{ Usage instruction }}} &
  \multicolumn{1}{l|}{\textbf{\rotatebox[origin=c]{90}{Access issues}}} &
  \multicolumn{1}{l|}{\textbf{\rotatebox[origin=c]{90}{Others}}} &
  \multicolumn{1}{l|}{{\color[HTML]{222222} \textbf{\rotatebox[origin=c]{90}{Compatibility}}}} &
  \multicolumn{1}{l|}{{\color[HTML]{222222} \textbf{\rotatebox[origin=c]{90}{Location}}}} &
  \multicolumn{1}{l|}{{\color[HTML]{222222} \textbf{\rotatebox[origin=c]{90}{Language}}}} &
  \multicolumn{1}{l|}{\textbf{\rotatebox[origin=c]{90}{Accessibility}}} &
  \multicolumn{1}{l|}{\textbf{\rotatebox[origin=c]{90}{Others}}} &
  \multicolumn{1}{l|}{\textbf{\rotatebox[origin=c]{90}{ Fulfilling interests }}} &
  \multicolumn{1}{l|}{{\color[HTML]{222222} \textbf{\rotatebox[origin=c]{90}{ Emotional aspects }}}} &
  \multicolumn{1}{l|}{{\color[HTML]{222222} \textbf{\rotatebox[origin=c]{90}{Preference}}}} &
  \multicolumn{1}{l|}{{\color[HTML]{222222} \textbf{\rotatebox[origin=c]{90}{Others}}}} &
  \multicolumn{1}{l|}{} \\ \cline{19-19} \hline
\multicolumn{1}{|l|}{\textbf{Firefox}} &
  \multicolumn{1}{c|}{6} &
  \multicolumn{1}{c|}{19} &
  \multicolumn{1}{c|}{23} &
  \multicolumn{1}{c|}{35} &
  \multicolumn{1}{c|}{9} &
  \multicolumn{1}{c|}{2} &
  \multicolumn{1}{c|}{0} &
  0 &
  \multicolumn{1}{c|}{1} &
  \multicolumn{1}{c|}{0} &
  \multicolumn{1}{c|}{0} &
  \multicolumn{1}{c|}{0} &
  0 &
  \multicolumn{1}{c|}{16} &
  \multicolumn{1}{c|}{3} &
  \multicolumn{1}{c|}{23} &
  0 & \multicolumn{1}{c|}{79} \\ \hline
\multicolumn{1}{|l|}{\textbf{K-9}} &
  \multicolumn{1}{c|}{2} &
  \multicolumn{1}{c|}{22} &
  \multicolumn{1}{c|}{14} &
  \multicolumn{1}{c|}{13} &
  \multicolumn{1}{c|}{0} &
  \multicolumn{1}{c|}{1} &
  \multicolumn{1}{c|}{1} &
  0 &
  \multicolumn{1}{c|}{5} &
  \multicolumn{1}{c|}{0} &
  \multicolumn{1}{c|}{0} &
  \multicolumn{1}{c|}{0} &
  0 &
  \multicolumn{1}{c|}{10} &
  \multicolumn{1}{c|}{1} &
  \multicolumn{1}{c|}{23} &
  0 & \multicolumn{1}{c|}{68} \\ \hline
\multicolumn{1}{|l|}{\textbf{Signal}} &
  \multicolumn{1}{c|}{8} &
  \multicolumn{1}{c|}{17} &
  \multicolumn{1}{c|}{4} &
  \multicolumn{1}{c|}{3} &
  \multicolumn{1}{c|}{5} &
  \multicolumn{1}{c|}{0} &
  \multicolumn{1}{c|}{0} &
  0 &
  \multicolumn{1}{c|}{0} &
  \multicolumn{1}{c|}{0} &
  \multicolumn{1}{c|}{0} &
  \multicolumn{1}{c|}{1} &
  0 &
  \multicolumn{1}{c|}{4} &
  \multicolumn{1}{c|}{0} &
  \multicolumn{1}{c|}{34} &
  0 & \multicolumn{1}{c|}{66} \\ \hline
\multicolumn{1}{|l|}{\textbf{Bitcoin}} &
  \multicolumn{1}{c|}{7} &
  \multicolumn{1}{c|}{12} &
  \multicolumn{1}{c|}{4} &
  \multicolumn{1}{c|}{5} &
  \multicolumn{1}{c|}{9} &
  \multicolumn{1}{c|}{5} &
  \multicolumn{1}{c|}{1} &
  10 &
  \multicolumn{1}{c|}{1} &
  \multicolumn{1}{c|}{1} &
  \multicolumn{1}{c|}{1} &
  \multicolumn{1}{c|}{1} &
  0 &
  \multicolumn{1}{c|}{5} &
  \multicolumn{1}{c|}{0} &
  \multicolumn{1}{c|}{7} &
  4 & \multicolumn{1}{c|}{58} \\ \hline
\multicolumn{1}{|l|}{\textbf{Osmand}} &
  \multicolumn{1}{c|}{1} &
  \multicolumn{1}{c|}{10} &
  \multicolumn{1}{c|}{4} &
  \multicolumn{1}{c|}{21} &
  \multicolumn{1}{c|}{0} &
  \multicolumn{1}{c|}{4} &
  \multicolumn{1}{c|}{0} &
  2 &
  \multicolumn{1}{c|}{0} &
  \multicolumn{1}{c|}{4} &
  \multicolumn{1}{c|}{1} &
  \multicolumn{1}{c|}{1} &
  0 &
  \multicolumn{1}{c|}{11} &
  \multicolumn{1}{c|}{0} &
  \multicolumn{1}{c|}{9} &
  0 & \multicolumn{1}{c|}{57} \\ \hline
\multicolumn{1}{|l|}{\textbf{Brave}} &
  \multicolumn{1}{c|}{1} &
  \multicolumn{1}{c|}{10} &
  \multicolumn{1}{c|}{12} &
  \multicolumn{1}{c|}{14} &
  \multicolumn{1}{c|}{2} &
  \multicolumn{1}{c|}{1} &
  \multicolumn{1}{c|}{0} &
  1 &
  \multicolumn{1}{c|}{0} &
  \multicolumn{1}{c|}{0} &
  \multicolumn{1}{c|}{0} &
  \multicolumn{1}{c|}{0} &
  1 &
  \multicolumn{1}{c|}{5} &
  \multicolumn{1}{c|}{2} &
  \multicolumn{1}{c|}{9} &
  0 & \multicolumn{1}{c|}{46} \\ \hline
\multicolumn{1}{|l|}{\textbf{Duckduckgo}} &
  \multicolumn{1}{c|}{4} &
  \multicolumn{1}{c|}{6} &
  \multicolumn{1}{c|}{3} &
  \multicolumn{1}{c|}{11} &
  \multicolumn{1}{c|}{3} &
  \multicolumn{1}{c|}{0} &
  \multicolumn{1}{c|}{2} &
  1 &
  \multicolumn{1}{c|}{1} &
  \multicolumn{1}{c|}{2} &
  \multicolumn{1}{c|}{1} &
  \multicolumn{1}{c|}{3} &
  1 &
  \multicolumn{1}{c|}{6} &
  \multicolumn{1}{c|}{1} &
  \multicolumn{1}{c|}{21} &
  0 & \multicolumn{1}{c|}{44} \\ \hline
\multicolumn{1}{|l|}{\textbf{Fbreader}} &
  \multicolumn{1}{c|}{2} &
  \multicolumn{1}{c|}{9} &
  \multicolumn{1}{c|}{9} &
  \multicolumn{1}{c|}{11} &
  \multicolumn{1}{c|}{0} &
  \multicolumn{1}{c|}{1} &
  \multicolumn{1}{c|}{0} &
  3 &
  \multicolumn{1}{c|}{1} &
  \multicolumn{1}{c|}{0} &
  \multicolumn{1}{c|}{0} &
  \multicolumn{1}{c|}{1} &
  1 &
  \multicolumn{1}{c|}{3} &
  \multicolumn{1}{c|}{1} &
  \multicolumn{1}{c|}{12} &
  0 & \multicolumn{1}{c|}{39} \\ \hline
\multicolumn{1}{|l|}{\textbf{Termux}} &
  \multicolumn{1}{c|}{1} &
  \multicolumn{1}{c|}{15} &
  \multicolumn{1}{c|}{2} &
  \multicolumn{1}{c|}{3} &
  \multicolumn{1}{c|}{2} &
  \multicolumn{1}{c|}{2} &
  \multicolumn{1}{c|}{4} &
  1 &
  \multicolumn{1}{c|}{1} &
  \multicolumn{1}{c|}{0} &
  \multicolumn{1}{c|}{0} &
  \multicolumn{1}{c|}{0} &
  0 &
  \multicolumn{1}{c|}{2} &
  \multicolumn{1}{c|}{0} &
  \multicolumn{1}{c|}{4} &
  0 & \multicolumn{1}{c|}{35} \\ \hline
\multicolumn{1}{|l|}{\textbf{Cgeo}} &
  \multicolumn{1}{c|}{1} &
  \multicolumn{1}{c|}{8} &
  \multicolumn{1}{c|}{2} &
  \multicolumn{1}{c|}{5} &
  \multicolumn{1}{c|}{0} &
  \multicolumn{1}{c|}{0} &
  \multicolumn{1}{c|}{5} &
  0 &
  \multicolumn{1}{c|}{0} &
  \multicolumn{1}{c|}{0} &
  \multicolumn{1}{c|}{0} &
  \multicolumn{1}{c|}{0} &
  0 &
  \multicolumn{1}{c|}{6} &
  \multicolumn{1}{c|}{0} &
  \multicolumn{1}{c|}{9} &
  0 & \multicolumn{1}{c|}{29} \\ \hline
\multicolumn{1}{|l|}{\textbf{WordPress}} &
  \multicolumn{1}{c|}{0} &
  \multicolumn{1}{c|}{10} &
  \multicolumn{1}{c|}{3} &
  \multicolumn{1}{c|}{6} &
  \multicolumn{1}{c|}{0} &
  \multicolumn{1}{c|}{3} &
  \multicolumn{1}{c|}{0} &
  1 &
  \multicolumn{1}{c|}{0} &
  \multicolumn{1}{c|}{0} &
  \multicolumn{1}{c|}{1} &
  \multicolumn{1}{c|}{0} &
  0 &
  \multicolumn{1}{c|}{1} &
  \multicolumn{1}{c|}{0} &
  \multicolumn{1}{c|}{8} &
  0 & \multicolumn{1}{c|}{28} \\ \hline
\multicolumn{1}{|l|}{\textbf{Pixel-dungeon}} &
  \multicolumn{1}{c|}{0} &
  \multicolumn{1}{c|}{2} &
  \multicolumn{1}{c|}{3} &
  \multicolumn{1}{c|}{4} &
  \multicolumn{1}{c|}{0} &
  \multicolumn{1}{c|}{0} &
  \multicolumn{1}{c|}{0} &
  1 &
  \multicolumn{1}{c|}{0} &
  \multicolumn{1}{c|}{0} &
  \multicolumn{1}{c|}{0} &
  \multicolumn{1}{c|}{0} &
  0 &
  \multicolumn{1}{c|}{1} &
  \multicolumn{1}{c|}{0} &
  \multicolumn{1}{c|}{8} &
  0 & \multicolumn{1}{c|}{17} \\ \hline
\multicolumn{1}{|l|}{\textbf{Total (\#)}} &
  \multicolumn{1}{c|}{33} &
  \multicolumn{1}{c|}{140} &
  \multicolumn{1}{c|}{83} &
  \multicolumn{1}{c|}{131} &
  \multicolumn{1}{c|}{30} &
  \multicolumn{1}{c|}{19} &
  \multicolumn{1}{c|}{13} &
  20 &
  \multicolumn{1}{c|}{10} &
  \multicolumn{1}{c|}{7} &
  \multicolumn{1}{c|}{4} &
  \multicolumn{1}{c|}{7} &
  3 &
  \multicolumn{1}{c|}{70} &
  \multicolumn{1}{c|}{8} &
  \multicolumn{1}{c|}{167} &
  4 & \\ \cline{1-18}
\multicolumn{1}{|l|}{\textbf{Total (\%)}} &
  \multicolumn{1}{c|}{2.75} &
  \multicolumn{1}{c|}{11.67} &
  \multicolumn{1}{c|}{6.92} &
  \multicolumn{1}{c|}{10.92} &
  \multicolumn{1}{c|}{2.50} &
  \multicolumn{1}{c|}{1.58} &
  \multicolumn{1}{c|}{1.08} &
  1.67 &
  \multicolumn{1}{c|}{0.83} &
  \multicolumn{1}{c|}{0.58} &
  \multicolumn{1}{c|}{0.33} &
  \multicolumn{1}{c|}{0.58} &
  0.25 &
  \multicolumn{1}{c|}{5.83} &
  \multicolumn{1}{c|}{0.67} &
  \multicolumn{1}{c|}{13.92} &
  0.33 & \\ \cline{1-18}
\end{tabular}
}
\end{table*}

\begin{figure}[h]
\centering
\includegraphics[scale=0.55]{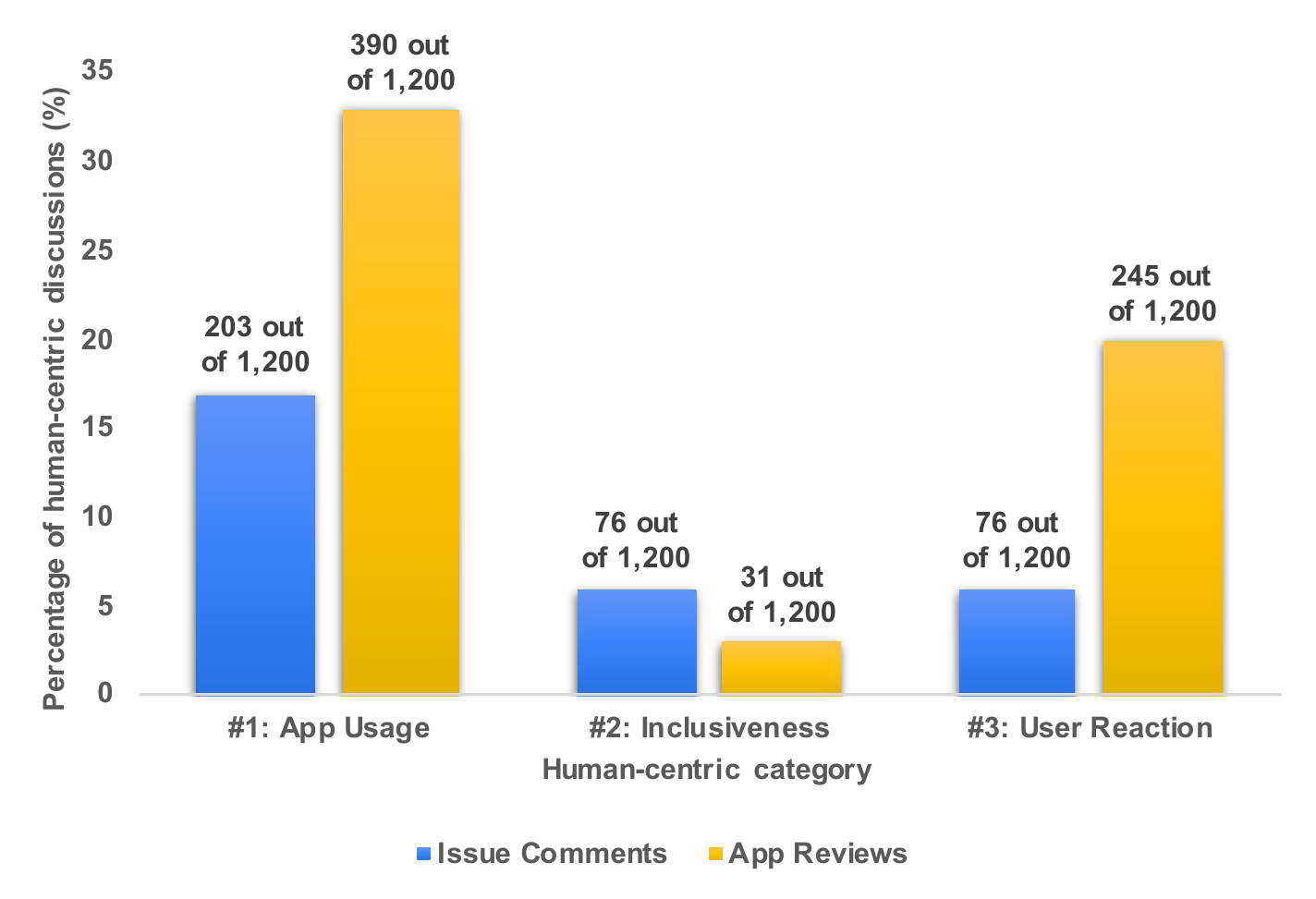}
\vspace{-3mm}
\caption{Number and percentage of human-centric issues out of 1,200 issue comments and 1,200 app reviews in 12 projects}
\label{fig:percentages}
\end{figure}

\subsection{Human-Centric Issues in Issue Comments Vs App Reviews and in Different Apps} \label{sec:categoriesProjects}

Table \ref{tbl:TableIssue} and \ref{tbl:TableApp} provide detailed information on the human-centric issues categories in the issue comments and app reviews of the 12 android projects. 
Overall, 25.5\% of the issue comments (306 out of 1,200 issue comments) and 47.25\% of the app reviews (567 out of 1,200 app reviews) studied in these 12 projects, discuss human-centric issues. These numbers suggest that human-centric issues are reported more frequently by the users through app reviews, than discussed by the developers through issue comments. How these human centric issues are spread among different categories, is shown in Figure \ref{fig:percentages}. 

In app reviews, out of the 567 human-centric issues, 390 reviews (32\% of total reviews) are related to App Usage, 31 (0.2\%) are related to Inclusiveness, and 245 (20\%) User Reactions related. On the other hand, among 306 human-centric developers discussions in GitHub issue comments, 203 reviews (17\% of total comments) are related to App Usage, 76 (6\%) related to Inclusiveness, and 76 (6\%) User Reactions related. Comments can belong to more than one category and therefore, the total percentage of different categories can be greater than the percentage of the total number of human-centric issues. Overall, App Usage followed by Inclusiveness and User Reaction are the most prevalence human-centric related discussion among developers, while App Usage, followed by User Reaction, and rarely Inclusiveness are more highly reported by the users.

The right columns in Table \ref{tbl:TableIssue} and \ref{tbl:TableApp} show the total number of issues/reviews out of the 100 selected ones that are having at least one human-centric issue, and can discuss various issues. In issue comments, Signal (35 out of 100) has the maximum number of issue comments discussing human-centric issues followed by Cgeo (33 out of 100) and Firefox (30 out of 100), while Pixel-dungeon (16 out of 100), FBreader (18 out of 100) and Bitcoin (19 out of 100) have the minimum number of discussions. 
In app reviews, Firefox (79 out of 100), followed by K-9 (68 out of 100) and Signal (66 out of 100) have the maximum number of human-centric issues reported by the users, and Pixel-dungeon (17 out of 100), has limited number of reported human-centric issues. 

  
\begin{center}
\begin{tcolorbox}[colback=white!2!white,colframe=black!75!black]
\textbf{RQ1 Summary.} Our taxonomy of human-centric issues constructed based on manual analysis of app reviews and issue comments include three high-level categories: \textbf{App Usage}, \textbf{Inclusiveness}, and \textbf{User Reaction}. Human-centric issues are reported more frequently by the users through app reviews, than being discussed by the developers in issue comments.  App Usage related issues are the most popular category to be discussed among both developers and end-users, followed by Inclusiveness among developers,  and User  Reaction by the users.
\end{tcolorbox}
\end{center}

\section{Automated Classification of End-user Human-centric Issues (RQ2)}\label{sec:RQ2appfindings}
\subsection{Approach}
In this section, we present our ML and DL models to automatically classify the app reviews and issue comments we collected and labeled, as described in Section \ref{sec:findings}. We only focused on identifying and classifying three high-level categories of human-centric issues in the taxonomy and adopted multi-label classification techniques for this purpose \cite{mao2005state}. 


\textbf{Dataset.}
We used the two datasets built in Section \ref{sec:findings}: 1) app reviews collected from Google Play Store, 2) issues comments from GitHub, and a combination of both app reviews and issues comments. Due to the data limitation, we aggregated the two datasets to see if merging the two varied sets of comments would help in model predictions. In total, we had a combination of 2,400 comments (1,200 app reviews, 1,200 issue comments) that we utilised to train our models for classification. We considered the classes as our three high-level categories identified in Section \ref{sec:findings}, and one extra category as a non-human-centric category. Non-human-centric refers to a comment/issue that does not belong to any of the human-centric categories.

\textbf{Method.}
We developed and applied various ML { and DL models} to our datasets and evaluated their performance based on various metrics. Before employing the models, we applied data pre-processing to ensure that the data is clean and free from any noise. Here we discuss the pre-processing steps and models in detail.

\subsubsection{Machine Learning}
\label{sec:ML}
\underline{\textbf{Pre-processing.}}
The pre-processing phase for our ML models spans over five major steps. 

\textit{Step 1. Convert Case:}
Conversion of all text to lower case helps in maintaining the consistency of the data, as it mitigates the same words of different cases being recounted in the vocabulary set. 

\textit{Step 2. Remove Noise:}
Noise usually affects the classification accuracy, training time and size of the classifier, which leads to a faulty or erroneous representation of the data. Noise handling technique involves \cite{gupta2019dealing}: 1) Removing numbers, 2) Removing punctuation and special characters, e.g.: “Hello!!!”, “PS: please reply me back”, and 3) Remove patterns, eg: “grreeetttttt”, “Soooon”. 

\textit{Step 3. Normalisation:}
Normalisation considers the abbreviations used and even though there is not any defined vocabulary for such words, some common occurrences of typically used words are converted to their original form. For example, “wasn’t” is converted to "was not".

\textit{Step 4. Stemming:}
The process of converting the word to its word stem is called stemming 
\cite{GeeksforGeeks}. 
We used the generally adopted Snowball stemmer, which is a better and aggressive version of porter stemmer \cite{GeeksforGeeks}.

\textit{Step 5. Stopwords:}
Words, such as “a”, “the”, “is”, “are”, which occur commonly and do not add valuable information to the modelling were removed in this step. 

\underline{\textbf{Feature Extraction.}}
In this step, 
we employed various techniques to extract features from the corpus. These techniques are detailed as follows. 

\textit{TF-IDF:}
In TF-IDF method, TF refers to term frequency in a document 
and IDF refers to inverse document frequency. TF-IDF gives the words occurring rarely more weight than the commonly occurring words. 
TF-IDF  supports functionalities such as analyser and ngrams to try and improve feature extraction techniques. We tried multiple ngrams such as (1,1)(1,2)(1,3)(2,2)(3,3)(4,4). Moreover, we tried switching the analyser to ‘char’ to extract features based on character n-gram combinations instead of words \cite{medium2019detailed}.    

\textit{Word2vec:}
Word2vec is a technique of processing text to numerical vectors which uses neural network model to create word embedding. We employed Google pre-trained word2vec model to convert words to their corresponding vector equivalent for a given sentence. 
This method helps in preserving the semantic of the sentence \cite{medium2019detailed}.

\textit{Word2vec \& TF-IDF stack:}
Employing word2vec and TF-IDF simultaneously helps get better performance \cite{datastack2018}. We investigated to know whether or not stacking the two models together will improve the feature extraction process. 

\underline{\textbf{Multi label classification strategies.}}
Some ML models are not built to classify multi label data and therefore we needed to apply one of the following methods to convert multi label data into multiple binary class problems \cite{prathibhamol2016multi}.

\textit{One-vs-Rest:}
It splits the data by accounting for one label at a time and grouping the rest. The process decomposes the labels into multiple binary classification problems where the model is equipped for each label. One-vs-Rest allows the model to focus on one attribute at a time, hence the model learns in a binary environment. We selected a sample of data from one class at a time and trained the binary classifier models to distinguish between the class and the others \cite{murphy2022probabilistic}. 

\textit{Classifier chains:}
It uses binary classification for multiple classes where each label is independently considered. This model works on interdependencies (correlation) of the label to boost performance and lower computational complexity. It is a feed forward model that takes input from the previous classifier and passes the output to the next one \cite{read2011classifier}.  

\underline{\textbf{Classification Models.}} We used several ML models suitable for multi-label classification.

\textit{Logistic regression:}
The Linear Logistic (LR) function is a powerful discriminative method for independent binary variables \cite{liu2014mlslr}. 
LR is found to be very efficient in performing multi-label classification 
\cite{prathibhamol2016multi}.

\textit{Support Vector Machine (SVM):}
Empirical study by Zhang et al. suggests that standalone SVM models are a viable option for multi-label classification \cite{li2006empirical}. 
We used a generic model, i.e. without Hyperboosting using ADTree or ADABoost techniques to check the baseline performance.

\textit{Random Forest (RF):}
RF is an ensemble method that uses multiple decision tree classifiers on chunks from a dataset. It averages the learning from each tree to improve its overall prediction. By using Gini index (joint score), RF can be used to simultaneously predict from multi label data \cite{kouchaki2020multi}.

\textit{XGBoost (XGB):}
This gradient boosting library implements ML algorithms under its framework. Using parallel tree boosting makes it highly efficient and flexible  
\cite{chen2019xgboost}. 

\begin{table*}[]
\centering
\caption{Results of different models on three different datasets}
\label{tbl:MLResults}
{\scriptsize
\renewcommand{\arraystretch}{1.5}

\begin{tabular}{cl|lllll|lllll|lllll|}
\cline{3-17}
\multicolumn{1}{l}{\cellcolor[HTML]{FFFFFF}\textit{}} &
  \textit{} &
  \multicolumn{5}{c|}{\cellcolor[HTML]{FFFFFF}{\textbf{App Reviews}}} &
  \multicolumn{5}{c|}{\cellcolor[HTML]{FFFFFF}{\textbf{Issue Comments}}} &
  \multicolumn{5}{c|}{\cellcolor[HTML]{FFFFFF}{\textbf{App reviews + Issue comments}}} \\ \hline
\rowcolor[HTML]{FFFFFF} 
\multicolumn{1}{|c|}{\cellcolor[HTML]{FFFFFF}{\textbf{\begin{tabular}[c]{@{}c@{}}ML \\ Strategy\end{tabular}}}} &
  \multicolumn{1}{c|}{\cellcolor[HTML]{FFFFFF}{\textbf{Models}}} &
  \multicolumn{1}{c|}{\cellcolor[HTML]{FFFFFF}{\textbf{\rotatebox[origin=c]{90}{Precision}}}} &
  \multicolumn{1}{c|}{\cellcolor[HTML]{FFFFFF}{\textbf{\rotatebox[origin=c]{90}{Recall}}}} &
  \multicolumn{1}{c|}{\cellcolor[HTML]{FFFFFF}{\textbf{\rotatebox[origin=c]{90}{Accuracy}}}} &
  \multicolumn{1}{c|}{\cellcolor[HTML]{FFFFFF}{\textbf{\rotatebox[origin=c]{90}{F1\_Score}}}} &
  \multicolumn{1}{c|}{\cellcolor[HTML]{FFFFFF}{\textbf{\rotatebox[origin=c]{90}{Hamming Loss}}}} &
  \multicolumn{1}{c|}{\cellcolor[HTML]{FFFFFF}{\textbf{\rotatebox[origin=c]{90}{Precision}}}} &
  \multicolumn{1}{c|}{\cellcolor[HTML]{FFFFFF}{\textbf{\rotatebox[origin=c]{90}{Recall}}}} &
  \multicolumn{1}{c|}{\cellcolor[HTML]{FFFFFF}{\textbf{\rotatebox[origin=c]{90}{Accuracy}}}} &
  \multicolumn{1}{c|}{\cellcolor[HTML]{FFFFFF}{\textbf{\rotatebox[origin=c]{90}{F1\_Score}}}} &
  \multicolumn{1}{c|}{\cellcolor[HTML]{FFFFFF}{\textbf{\rotatebox[origin=c]{90}{Hamming Loss}}}} &
  \multicolumn{1}{c|}{\cellcolor[HTML]{FFFFFF}{\textbf{\rotatebox[origin=c]{90}{Precision}}}} &
  \multicolumn{1}{c|}{\cellcolor[HTML]{FFFFFF}{\textbf{\rotatebox[origin=c]{90}{Recall}}}} &
  \multicolumn{1}{c|}{\cellcolor[HTML]{FFFFFF}{\textbf{\rotatebox[origin=c]{90}{Accuracy}}}} &
  \multicolumn{1}{c|}{\cellcolor[HTML]{FFFFFF}{\textbf{\rotatebox[origin=c]{90}{F1\_Score}}}} &
  \multicolumn{1}{c|}{\cellcolor[HTML]{FFFFFF}{\textbf{\rotatebox[origin=c]{90}{Hamming Loss}}}} \\ \hline
\rowcolor[HTML]{FFFFFF} 
\multicolumn{1}{|c|}{\cellcolor[HTML]{FFFFFF}} &
  {tf–idf + LR} &
  \multicolumn{1}{l|}{\cellcolor[HTML]{FFFFFF}{0.73}} &
  \multicolumn{1}{l|}{\cellcolor[HTML]{FFFFFF}{0.71}} &
  \multicolumn{1}{l|}{\cellcolor[HTML]{FFFFFF}{0.61}} &
  \multicolumn{1}{l|}{\cellcolor[HTML]{FFFFFF}{0.72}} &
  {0.15} &
  \multicolumn{1}{l|}{\cellcolor[HTML]{FFFFFF}{0.83}} &
  \multicolumn{1}{l|}{\cellcolor[HTML]{FFFFFF}{0.75}} &
  \multicolumn{1}{l|}{\cellcolor[HTML]{FFFFFF}{0.73}} &
  \multicolumn{1}{l|}{\cellcolor[HTML]{FFFFFF}{0.79}} &
  {0.11} &
  \multicolumn{1}{l|}{\cellcolor[HTML]{FFFFFF}{0.76}} &
  \multicolumn{1}{l|}{\cellcolor[HTML]{FFFFFF}{0.74}} &
  \multicolumn{1}{l|}{\cellcolor[HTML]{FFFFFF}{0.67}} &
  \multicolumn{1}{l|}{\cellcolor[HTML]{FFFFFF}{0.75}} &
  {0.13} \\ \cline{2-17} 
\rowcolor[HTML]{FFFFFF} 
\multicolumn{1}{|c|}{\cellcolor[HTML]{FFFFFF}} &
  {tf–idf + SVM} &
  \multicolumn{1}{l|}{\cellcolor[HTML]{FFFFFF}{0.75}} &
  \multicolumn{1}{l|}{\cellcolor[HTML]{FFFFFF}{0.69}} &
  \multicolumn{1}{l|}{\cellcolor[HTML]{FFFFFF}{0.60}} &
  \multicolumn{1}{l|}{\cellcolor[HTML]{FFFFFF}{0.72}} &
  {0.15} &
  \multicolumn{1}{l|}{\cellcolor[HTML]{FFFFFF}{0.82}} &
  \multicolumn{1}{l|}{\cellcolor[HTML]{FFFFFF}{0.70}} &
  \multicolumn{1}{l|}{\cellcolor[HTML]{FFFFFF}{0.72}} &
  \multicolumn{1}{l|}{\cellcolor[HTML]{FFFFFF}{0.75}} &
  {0.12} &
  \multicolumn{1}{l|}{\cellcolor[HTML]{FFFFFF}{0.81}} &
  \multicolumn{1}{l|}{\cellcolor[HTML]{FFFFFF}{0.60}} &
  \multicolumn{1}{l|}{\cellcolor[HTML]{FFFFFF}{0.62}} &
  \multicolumn{1}{l|}{\cellcolor[HTML]{FFFFFF}{0.69}} &
  {0.14} \\ \cline{2-17} 
\rowcolor[HTML]{FFFFFF} 
\multicolumn{1}{|c|}{\cellcolor[HTML]{FFFFFF}} &
  {tf–idf + RF} &
  \multicolumn{1}{l|}{\cellcolor[HTML]{FFFFFF}{0.77}} &
  \multicolumn{1}{l|}{\cellcolor[HTML]{FFFFFF}{0.58}} &
  \multicolumn{1}{l|}{\cellcolor[HTML]{FFFFFF}{0.56}} &
  \multicolumn{1}{l|}{\cellcolor[HTML]{FFFFFF}{0.66}} &
  {0.16} &
  \multicolumn{1}{l|}{\cellcolor[HTML]{FFFFFF}{0.83}} &
  \multicolumn{1}{l|}{\cellcolor[HTML]{FFFFFF}{0.70}} &
  \multicolumn{1}{l|}{\cellcolor[HTML]{FFFFFF}{0.71}} &
  \multicolumn{1}{l|}{\cellcolor[HTML]{FFFFFF}{0.76}} &
  {0.11} &
  \multicolumn{1}{l|}{\cellcolor[HTML]{FFFFFF}{0.78}} &
  \multicolumn{1}{l|}{\cellcolor[HTML]{FFFFFF}{0.58}} &
  \multicolumn{1}{l|}{\cellcolor[HTML]{FFFFFF}{0.60}} &
  \multicolumn{1}{l|}{\cellcolor[HTML]{FFFFFF}{0.67}} &
  {0.15} \\ \cline{2-17} 
\rowcolor[HTML]{FFFFFF} 
\multicolumn{1}{|c|}{\cellcolor[HTML]{FFFFFF}} &
  {tf–idf + XGB} &
  \multicolumn{1}{l|}{\cellcolor[HTML]{FFFFFF}{0.71}} &
  \multicolumn{1}{l|}{\cellcolor[HTML]{FFFFFF}{0.62}} &
  \multicolumn{1}{l|}{\cellcolor[HTML]{FFFFFF}{0.55}} &
  \multicolumn{1}{l|}{\cellcolor[HTML]{FFFFFF}{0.66}} &
  {0.18} &
  \multicolumn{1}{l|}{\cellcolor[HTML]{FFFFFF}{0.81}} &
  \multicolumn{1}{l|}{\cellcolor[HTML]{FFFFFF}{0.71}} &
  \multicolumn{1}{l|}{\cellcolor[HTML]{FFFFFF}{0.69}} &
  \multicolumn{1}{l|}{\cellcolor[HTML]{FFFFFF}{0.75}} &
  {0.12} &
  \multicolumn{1}{l|}{\cellcolor[HTML]{FFFFFF}{0.77}} &
  \multicolumn{1}{l|}{\cellcolor[HTML]{FFFFFF}{0.66}} &
  \multicolumn{1}{l|}{\cellcolor[HTML]{FFFFFF}{0.63}} &
  \multicolumn{1}{l|}{\cellcolor[HTML]{FFFFFF}{0.71}} &
  {0.14} \\ \cline{2-17} 
\rowcolor[HTML]{FFFFFF} 
\multicolumn{1}{|c|}{\cellcolor[HTML]{FFFFFF}} &
  {word2vec + LR} &
  \multicolumn{1}{l|}{\cellcolor[HTML]{FFFFFF}{0.58}} &
  \multicolumn{1}{l|}{\cellcolor[HTML]{FFFFFF}{0.73}} &
  \multicolumn{1}{l|}{\cellcolor[HTML]{FFFFFF}{0.47}} &
  \multicolumn{1}{l|}{\cellcolor[HTML]{FFFFFF}{0.65}} &
  {0.22} &
  \multicolumn{1}{l|}{\cellcolor[HTML]{FFFFFF}{0.58}} &
  \multicolumn{1}{l|}{\cellcolor[HTML]{FFFFFF}{0.73}} &
  \multicolumn{1}{l|}{\cellcolor[HTML]{FFFFFF}{0.47}} &
  \multicolumn{1}{l|}{\cellcolor[HTML]{FFFFFF}{0.65}} &
  {0.22} &
  \multicolumn{1}{l|}{\cellcolor[HTML]{FFFFFF}{0.50}} &
  \multicolumn{1}{l|}{\cellcolor[HTML]{FFFFFF}{0.71}} &
  \multicolumn{1}{l|}{\cellcolor[HTML]{FFFFFF}{0.39}} &
  \multicolumn{1}{l|}{\cellcolor[HTML]{FFFFFF}{0.58}} &
  {0.27} \\ \cline{2-17} 
\rowcolor[HTML]{FFFFFF} 
\multicolumn{1}{|c|}{\cellcolor[HTML]{FFFFFF}} &
  {word2vec + SVM} &
  \multicolumn{1}{l|}{\cellcolor[HTML]{FFFFFF}{0.57}} &
  \multicolumn{1}{l|}{\cellcolor[HTML]{FFFFFF}{0.73}} &
  \multicolumn{1}{l|}{\cellcolor[HTML]{FFFFFF}{0.45}} &
  \multicolumn{1}{l|}{\cellcolor[HTML]{FFFFFF}{0.64}} &
  {0.23} &
  \multicolumn{1}{l|}{\cellcolor[HTML]{FFFFFF}{0.79}} &
  \multicolumn{1}{l|}{\cellcolor[HTML]{FFFFFF}{0.69}} &
  \multicolumn{1}{l|}{\cellcolor[HTML]{FFFFFF}{0.71}} &
  \multicolumn{1}{l|}{\cellcolor[HTML]{FFFFFF}{0.73}} &
  {0.13} &
  \multicolumn{1}{l|}{\cellcolor[HTML]{FFFFFF}{0.82}} &
  \multicolumn{1}{l|}{\cellcolor[HTML]{FFFFFF}{0.58}} &
  \multicolumn{1}{l|}{\cellcolor[HTML]{FFFFFF}{0.60}} &
  \multicolumn{1}{l|}{\cellcolor[HTML]{FFFFFF}{0.68}} &
  {0.15} \\ \cline{2-17} 
\rowcolor[HTML]{FFFFFF} 
\multicolumn{1}{|c|}{\cellcolor[HTML]{FFFFFF}} &
  {word2vec + RF} &
  \multicolumn{1}{l|}{\cellcolor[HTML]{FFFFFF}{0.80}} &
  \multicolumn{1}{l|}{\cellcolor[HTML]{FFFFFF}{0.50}} &
  \multicolumn{1}{l|}{\cellcolor[HTML]{FFFFFF}{0.49}} &
  \multicolumn{1}{l|}{\cellcolor[HTML]{FFFFFF}{0.61}} &
  {0.17} &
  \multicolumn{1}{l|}{\cellcolor[HTML]{FFFFFF}{0.78}} &
  \multicolumn{1}{l|}{\cellcolor[HTML]{FFFFFF}{0.74}} &
  \multicolumn{1}{l|}{\cellcolor[HTML]{FFFFFF}{0.74}} &
  \multicolumn{1}{l|}{\cellcolor[HTML]{FFFFFF}{0.76}} &
  {0.12} &
  \multicolumn{1}{l|}{\cellcolor[HTML]{FFFFFF}{0.75}} &
  \multicolumn{1}{l|}{\cellcolor[HTML]{FFFFFF}{0.62}} &
  \multicolumn{1}{l|}{\cellcolor[HTML]{FFFFFF}{0.63}} &
  \multicolumn{1}{l|}{\cellcolor[HTML]{FFFFFF}{0.68}} &
  {0.16} \\ \cline{2-17} 
\rowcolor[HTML]{FFFFFF} 
\multicolumn{1}{|c|}{\cellcolor[HTML]{FFFFFF}} &
  {word2vec + XGB} &
  \multicolumn{1}{l|}{\cellcolor[HTML]{FFFFFF}{0.77}} &
  \multicolumn{1}{l|}{\cellcolor[HTML]{FFFFFF}{0.57}} &
  \multicolumn{1}{l|}{\cellcolor[HTML]{FFFFFF}{0.53}} &
  \multicolumn{1}{l|}{\cellcolor[HTML]{FFFFFF}{0.66}} &
  {0.17} &
  \multicolumn{1}{l|}{\cellcolor[HTML]{FFFFFF}{0.79}} &
  \multicolumn{1}{l|}{\cellcolor[HTML]{FFFFFF}{0.69}} &
  \multicolumn{1}{l|}{\cellcolor[HTML]{FFFFFF}{0.71}} &
  \multicolumn{1}{l|}{\cellcolor[HTML]{FFFFFF}{0.73}} &
  {0.13} &
  \multicolumn{1}{l|}{\cellcolor[HTML]{FFFFFF}{0.77}} &
  \multicolumn{1}{l|}{\cellcolor[HTML]{FFFFFF}{0.61}} &
  \multicolumn{1}{l|}{\cellcolor[HTML]{FFFFFF}{0.60}} &
  \multicolumn{1}{l|}{\cellcolor[HTML]{FFFFFF}{0.68}} &
  {0.15} \\ \cline{2-17} 
\rowcolor[HTML]{FFFFFF} 
\multicolumn{1}{|c|}{\cellcolor[HTML]{FFFFFF}} &
  {word2vec + tf–idf + LR} &
  \multicolumn{1}{l|}{\cellcolor[HTML]{FFFFFF}{0.68}} &
  \multicolumn{1}{l|}{\cellcolor[HTML]{FFFFFF}{0.73}} &
  \multicolumn{1}{l|}{\cellcolor[HTML]{FFFFFF}{0.57}} &
  \multicolumn{1}{l|}{\cellcolor[HTML]{FFFFFF}{0.70}} &
  {0.17} &
  \multicolumn{1}{l|}{\cellcolor[HTML]{FFFFFF}{0.73}} &
  \multicolumn{1}{l|}{\cellcolor[HTML]{FFFFFF}{0.74}} &
  \multicolumn{1}{l|}{\cellcolor[HTML]{FFFFFF}{0.67}} &
  \multicolumn{1}{l|}{\cellcolor[HTML]{FFFFFF}{0.73}} &
  {0.14} &
  \multicolumn{1}{l|}{\cellcolor[HTML]{FFFFFF}{0.69}} &
  \multicolumn{1}{l|}{\cellcolor[HTML]{FFFFFF}{0.74}} &
  \multicolumn{1}{l|}{\cellcolor[HTML]{FFFFFF}{0.60}} &
  \multicolumn{1}{l|}{\cellcolor[HTML]{FFFFFF}{0.71}} &
  {0.16} \\ \cline{2-17} 
\rowcolor[HTML]{FFFFFF} 
\multicolumn{1}{|c|}{\cellcolor[HTML]{FFFFFF}} &
  {word2vec + tf–idf + SVM} &
  \multicolumn{1}{l|}{\cellcolor[HTML]{FFFFFF}{0.69}} &
  \multicolumn{1}{l|}{\cellcolor[HTML]{FFFFFF}{0.69}} &
  \multicolumn{1}{l|}{\cellcolor[HTML]{FFFFFF}{0.54}} &
  \multicolumn{1}{l|}{\cellcolor[HTML]{FFFFFF}{0.69}} &
  {0.17} &
  \multicolumn{1}{l|}{\cellcolor[HTML]{FFFFFF}{0.81}} &
  \multicolumn{1}{l|}{\cellcolor[HTML]{FFFFFF}{0.72}} &
  \multicolumn{1}{l|}{\cellcolor[HTML]{FFFFFF}{0.74}} &
  \multicolumn{1}{l|}{\cellcolor[HTML]{FFFFFF}{0.76}} &
  {0.12} &
  \multicolumn{1}{l|}{\cellcolor[HTML]{FFFFFF}{0.84}} &
  \multicolumn{1}{l|}{\cellcolor[HTML]{FFFFFF}{0.62}} &
  \multicolumn{1}{l|}{\cellcolor[HTML]{FFFFFF}{0.63}} &
  \multicolumn{1}{l|}{\cellcolor[HTML]{FFFFFF}{0.71}} &
  {0.13} \\ \cline{2-17} 
\rowcolor[HTML]{FFFFFF} 
\multicolumn{1}{|c|}{\cellcolor[HTML]{FFFFFF}} &
  {word2vec + tf–idf + RF} &
  \multicolumn{1}{l|}{\cellcolor[HTML]{FFFFFF}{0.81}} &
  \multicolumn{1}{l|}{\cellcolor[HTML]{FFFFFF}{0.54}} &
  \multicolumn{1}{l|}{\cellcolor[HTML]{FFFFFF}{0.54}} &
  \multicolumn{1}{l|}{\cellcolor[HTML]{FFFFFF}{0.65}} &
  {0.16} &
  \multicolumn{1}{l|}{\cellcolor[HTML]{FFFFFF}{0.79}} &
  \multicolumn{1}{l|}{\cellcolor[HTML]{FFFFFF}{0.73}} &
  \multicolumn{1}{l|}{\cellcolor[HTML]{FFFFFF}{0.74}} &
  \multicolumn{1}{l|}{\cellcolor[HTML]{FFFFFF}{0.76}} &
  {0.12} &
  \multicolumn{1}{l|}{\cellcolor[HTML]{FFFFFF}{0.77}} &
  \multicolumn{1}{l|}{\cellcolor[HTML]{FFFFFF}{0.63}} &
  \multicolumn{1}{l|}{\cellcolor[HTML]{FFFFFF}{0.64}} &
  \multicolumn{1}{l|}{\cellcolor[HTML]{FFFFFF}{0.69}} &
  {0.15} \\ \cline{2-17}
\rowcolor[HTML]{FFFFFF} 
\multicolumn{1}{|c|}{\multirow{-12}{*}{\cellcolor[HTML]{FFFFFF}{\textbf{\rotatebox[origin=c]{90}{One vs Rest}}}}} &
  {word2vec + tf–idf + XGB} &
  \multicolumn{1}{l|}{\cellcolor[HTML]{FFFFFF}{0.78}} &
  \multicolumn{1}{l|}{\cellcolor[HTML]{FFFFFF}{0.61}} &
  \multicolumn{1}{l|}{\cellcolor[HTML]{FFFFFF}{0.55}} &
  \multicolumn{1}{l|}{\cellcolor[HTML]{FFFFFF}{0.68}} &
  {0.15} &
  \multicolumn{1}{l|}{\cellcolor[HTML]{FFFFFF}{0.83}} &
  \multicolumn{1}{l|}{\cellcolor[HTML]{FFFFFF}{0.72}} &
  \multicolumn{1}{l|}{\cellcolor[HTML]{FFFFFF}{0.71}} &
  \multicolumn{1}{l|}{\cellcolor[HTML]{FFFFFF}{0.77}} &
  {0.11} &
  \multicolumn{1}{l|}{\cellcolor[HTML]{FFFFFF}{0.80}} &
  \multicolumn{1}{l|}{\cellcolor[HTML]{FFFFFF}{0.64}} &
  \multicolumn{1}{l|}{\cellcolor[HTML]{FFFFFF}{0.62}} &
  \multicolumn{1}{l|}{\cellcolor[HTML]{FFFFFF}{0.71}} &
  {0.14} \\ \hline
\rowcolor[HTML]{FFFFFF} 
\multicolumn{1}{|c|}{\cellcolor[HTML]{FFFFFF}} &
  {tf–idf + LR} &
  \multicolumn{1}{l|}{\cellcolor[HTML]{FFFFFF}{0.73}} &
  \multicolumn{1}{l|}{\cellcolor[HTML]{FFFFFF}{0.72}} &
  \multicolumn{1}{l|}{\cellcolor[HTML]{FFFFFF}{0.66}} &
  \multicolumn{1}{l|}{\cellcolor[HTML]{FFFFFF}{0.73}} &
  {0.15} &
  \multicolumn{1}{l|}{\cellcolor[HTML]{FFFFFF}{0.77}} &
  \multicolumn{1}{l|}{\cellcolor[HTML]{FFFFFF}{0.81}} &
  \multicolumn{1}{l|}{\cellcolor[HTML]{FFFFFF}{0.76}} &
  \multicolumn{1}{l|}{\cellcolor[HTML]{FFFFFF}{0.79}} &
  {0.11} &
  \multicolumn{1}{l|}{\cellcolor[HTML]{FFFFFF}{0.74}} &
  \multicolumn{1}{l|}{\cellcolor[HTML]{FFFFFF}{0.76}} &
  \multicolumn{1}{l|}{\cellcolor[HTML]{FFFFFF}{0.70}} &
  \multicolumn{1}{l|}{\cellcolor[HTML]{FFFFFF}{0.75}} &
  {0.13} \\ \cline{2-17} 
\rowcolor[HTML]{FFFFFF} 
\multicolumn{1}{|c|}{\cellcolor[HTML]{FFFFFF}} &
  {tf–idf + SVM} &
  \multicolumn{1}{l|}{\cellcolor[HTML]{FFFFFF}{0.74}} &
  \multicolumn{1}{l|}{\cellcolor[HTML]{FFFFFF}{0.71}} &
  \multicolumn{1}{l|}{\cellcolor[HTML]{FFFFFF}{0.65}} &
  \multicolumn{1}{l|}{\cellcolor[HTML]{FFFFFF}{0.72}} &
  {0.15} &
  \multicolumn{1}{l|}{\cellcolor[HTML]{FFFFFF}{0.80}} &
  \multicolumn{1}{l|}{\cellcolor[HTML]{FFFFFF}{0.79}} &
  \multicolumn{1}{l|}{\cellcolor[HTML]{FFFFFF}{0.79}} &
  \multicolumn{1}{l|}{\cellcolor[HTML]{FFFFFF}{0.80}} &
  {0.10} &
  \multicolumn{1}{l|}{\cellcolor[HTML]{FFFFFF}{0.76}} &
  \multicolumn{1}{l|}{\cellcolor[HTML]{FFFFFF}{0.74}} &
  \multicolumn{1}{l|}{\cellcolor[HTML]{FFFFFF}{0.72}} &
  \multicolumn{1}{l|}{\cellcolor[HTML]{FFFFFF}{0.75}} &
  {0.13} \\ \cline{2-17} 
\rowcolor[HTML]{FFFFFF} 
\multicolumn{1}{|c|}{\cellcolor[HTML]{FFFFFF}} &
  {tf–idf + RF} &
  \multicolumn{1}{l|}{\cellcolor[HTML]{FFFFFF}{0.75}} &
  \multicolumn{1}{l|}{\cellcolor[HTML]{FFFFFF}{0.62}} &
  \multicolumn{1}{l|}{\cellcolor[HTML]{FFFFFF}{0.60}} &
  \multicolumn{1}{l|}{\cellcolor[HTML]{FFFFFF}{0.68}} &
  {0.16} &
  \multicolumn{1}{l|}{\cellcolor[HTML]{FFFFFF}{0.81}} &
  \multicolumn{1}{l|}{\cellcolor[HTML]{FFFFFF}{0.70}} &
  \multicolumn{1}{l|}{\cellcolor[HTML]{FFFFFF}{0.72}} &
  \multicolumn{1}{l|}{\cellcolor[HTML]{FFFFFF}{0.75}} &
  {0.12} &
  \multicolumn{1}{l|}{\cellcolor[HTML]{FFFFFF}{0.75}} &
  \multicolumn{1}{l|}{\cellcolor[HTML]{FFFFFF}{0.62}} &
  \multicolumn{1}{l|}{\cellcolor[HTML]{FFFFFF}{0.65}} &
  \multicolumn{1}{l|}{\cellcolor[HTML]{FFFFFF}{0.68}} &
  {0.16} \\ \cline{2-17} 
\rowcolor[HTML]{FFFFFF} 
\multicolumn{1}{|c|}{\cellcolor[HTML]{FFFFFF}} &
  {tf–idf + XGB} &
  \multicolumn{1}{l|}{\cellcolor[HTML]{FFFFFF}{0.68}} &
  \multicolumn{1}{l|}{\cellcolor[HTML]{FFFFFF}{0.65}} &
  \multicolumn{1}{l|}{\cellcolor[HTML]{FFFFFF}{0.61}} &
  \multicolumn{1}{l|}{\cellcolor[HTML]{FFFFFF}{0.66}} &
  {0.18} &
  \multicolumn{1}{l|}{\cellcolor[HTML]{FFFFFF}{0.78}} &
  \multicolumn{1}{l|}{\cellcolor[HTML]{FFFFFF}{0.76}} &
  \multicolumn{1}{l|}{\cellcolor[HTML]{FFFFFF}{0.76}} &
  \multicolumn{1}{l|}{\cellcolor[HTML]{FFFFFF}{0.77}} &
  {0.12} &
  \multicolumn{1}{l|}{\cellcolor[HTML]{FFFFFF}{0.72}} &
  \multicolumn{1}{l|}{\cellcolor[HTML]{FFFFFF}{0.70}} &
  \multicolumn{1}{l|}{\cellcolor[HTML]{FFFFFF}{0.69}} &
  \multicolumn{1}{l|}{\cellcolor[HTML]{FFFFFF}{0.71}} &
  {0.15} \\ \cline{2-17} 
\rowcolor[HTML]{FFFFFF} 
\multicolumn{1}{|c|}{\cellcolor[HTML]{FFFFFF}} &
  {word2vec + LR} &
  \multicolumn{1}{l|}{\cellcolor[HTML]{FFFFFF}{0.59}} &
  \multicolumn{1}{l|}{\cellcolor[HTML]{FFFFFF}{0.69}} &
  \multicolumn{1}{l|}{\cellcolor[HTML]{FFFFFF}{0.52}} &
  \multicolumn{1}{l|}{\cellcolor[HTML]{FFFFFF}{0.64}} &
  {0.22} &
  \multicolumn{1}{l|}{\cellcolor[HTML]{FFFFFF}{0.38}} &
  \multicolumn{1}{l|}{\cellcolor[HTML]{FFFFFF}{0.54}} &
  \multicolumn{1}{l|}{\cellcolor[HTML]{FFFFFF}{0.44}} &
  \multicolumn{1}{l|}{\cellcolor[HTML]{FFFFFF}{0.45}} &
  {0.34} &
  \multicolumn{1}{l|}{\cellcolor[HTML]{FFFFFF}{0.44}} &
  \multicolumn{1}{l|}{\cellcolor[HTML]{FFFFFF}{0.57}} &
  \multicolumn{1}{l|}{\cellcolor[HTML]{FFFFFF}{0.41}} &
  \multicolumn{1}{l|}{\cellcolor[HTML]{FFFFFF}{0.50}} &
  {0.31} \\ \cline{2-17} 
\rowcolor[HTML]{FFFFFF} 
\multicolumn{1}{|c|}{\cellcolor[HTML]{FFFFFF}} &
  {word2vec + SVM} &
  \multicolumn{1}{l|}{\cellcolor[HTML]{FFFFFF}{0.57}} &
  \multicolumn{1}{l|}{\cellcolor[HTML]{FFFFFF}{0.65}} &
  \multicolumn{1}{l|}{\cellcolor[HTML]{FFFFFF}{0.50}} &
  \multicolumn{1}{l|}{\cellcolor[HTML]{FFFFFF}{0.61}} &
  {0.23} &
  \multicolumn{1}{l|}{\cellcolor[HTML]{FFFFFF}{0.74}} &
  \multicolumn{1}{l|}{\cellcolor[HTML]{FFFFFF}{0.71}} &
  \multicolumn{1}{l|}{\cellcolor[HTML]{FFFFFF}{0.74}} &
  \multicolumn{1}{l|}{\cellcolor[HTML]{FFFFFF}{0.73}} &
  {0.14} &
  \multicolumn{1}{l|}{\cellcolor[HTML]{FFFFFF}{0.69}} &
  \multicolumn{1}{l|}{\cellcolor[HTML]{FFFFFF}{0.65}} &
  \multicolumn{1}{l|}{\cellcolor[HTML]{FFFFFF}{0.67}} &
  \multicolumn{1}{l|}{\cellcolor[HTML]{FFFFFF}{0.67}} &
  {0.17} \\ \cline{2-17} 
\rowcolor[HTML]{FFFFFF} 
\multicolumn{1}{|c|}{\cellcolor[HTML]{FFFFFF}} &
  {word2vec + RF} &
  \multicolumn{1}{l|}{\cellcolor[HTML]{FFFFFF}{0.70}} &
  \multicolumn{1}{l|}{\cellcolor[HTML]{FFFFFF}{0.57}} &
  \multicolumn{1}{l|}{\cellcolor[HTML]{FFFFFF}{0.57}} &
  \multicolumn{1}{l|}{\cellcolor[HTML]{FFFFFF}{0.63}} &
  {0.19} &
  \multicolumn{1}{l|}{\cellcolor[HTML]{FFFFFF}{0.76}} &
  \multicolumn{1}{l|}{\cellcolor[HTML]{FFFFFF}{0.75}} &
  \multicolumn{1}{l|}{\cellcolor[HTML]{FFFFFF}{0.75}} &
  \multicolumn{1}{l|}{\cellcolor[HTML]{FFFFFF}{0.76}} &
  {0.13} &
  \multicolumn{1}{l|}{\cellcolor[HTML]{FFFFFF}{0.70}} &
  \multicolumn{1}{l|}{\cellcolor[HTML]{FFFFFF}{0.67}} &
  \multicolumn{1}{l|}{\cellcolor[HTML]{FFFFFF}{0.68}} &
  \multicolumn{1}{l|}{\cellcolor[HTML]{FFFFFF}{0.68}} &
  {0.17} \\ \cline{2-17} 
\rowcolor[HTML]{FFFFFF} 
\multicolumn{1}{|c|}{\cellcolor[HTML]{FFFFFF}} &
  {word2vec + XGB} &
  \multicolumn{1}{l|}{\cellcolor[HTML]{FFFFFF}{0.69}} &
  \multicolumn{1}{l|}{\cellcolor[HTML]{FFFFFF}{0.65}} &
  \multicolumn{1}{l|}{\cellcolor[HTML]{FFFFFF}{0.63}} &
  \multicolumn{1}{l|}{\cellcolor[HTML]{FFFFFF}{0.67}} &
  {0.18} &
  \multicolumn{1}{l|}{\cellcolor[HTML]{FFFFFF}{0.78}} &
  \multicolumn{1}{l|}{\cellcolor[HTML]{FFFFFF}{0.76}} &
  \multicolumn{1}{l|}{\cellcolor[HTML]{FFFFFF}{0.77}} &
  \multicolumn{1}{l|}{\cellcolor[HTML]{FFFFFF}{0.77}} &
  {0.12} &
  \multicolumn{1}{l|}{\cellcolor[HTML]{FFFFFF}{0.71}} &
  \multicolumn{1}{l|}{\cellcolor[HTML]{FFFFFF}{0.67}} &
  \multicolumn{1}{l|}{\cellcolor[HTML]{FFFFFF}{0.68}} &
  \multicolumn{1}{l|}{\cellcolor[HTML]{FFFFFF}{0.69}} &
  {0.16} \\ \cline{2-17} 
\rowcolor[HTML]{FFFFFF} 
\multicolumn{1}{|c|}{\cellcolor[HTML]{FFFFFF}} &
  {word2vec + tf–idf + LR} &
  \multicolumn{1}{l|}{\cellcolor[HTML]{FFFFFF}{0.68}} &
  \multicolumn{1}{l|}{\cellcolor[HTML]{FFFFFF}{0.71}} &
  \multicolumn{1}{l|}{\cellcolor[HTML]{FFFFFF}{0.59}} &
  \multicolumn{1}{l|}{\cellcolor[HTML]{FFFFFF}{0.70}} &
  {0.17} &
  \multicolumn{1}{l|}{\cellcolor[HTML]{FFFFFF}{0.69}} &
  \multicolumn{1}{l|}{\cellcolor[HTML]{FFFFFF}{0.76}} &
  \multicolumn{1}{l|}{\cellcolor[HTML]{FFFFFF}{0.70}} &
  \multicolumn{1}{l|}{\cellcolor[HTML]{FFFFFF}{0.72}} &
  {0.15} &
  \multicolumn{1}{l|}{\cellcolor[HTML]{FFFFFF}{0.69}} &
  \multicolumn{1}{l|}{\cellcolor[HTML]{FFFFFF}{0.74}} &
  \multicolumn{1}{l|}{\cellcolor[HTML]{FFFFFF}{0.65}} &
  \multicolumn{1}{l|}{\cellcolor[HTML]{FFFFFF}{0.71}} &
  {0.16} \\ \cline{2-17} 
\rowcolor[HTML]{FFFFFF} 
\multicolumn{1}{|c|}{\cellcolor[HTML]{FFFFFF}} &
  {word2vec + tf–idf + SVM} &
  \multicolumn{1}{l|}{\cellcolor[HTML]{FFFFFF}{0.69}} &
  \multicolumn{1}{l|}{\cellcolor[HTML]{FFFFFF}{0.68}} &
  \multicolumn{1}{l|}{\cellcolor[HTML]{FFFFFF}{0.62}} &
  \multicolumn{1}{l|}{\cellcolor[HTML]{FFFFFF}{0.68}} &
  {0.17} &
  \multicolumn{1}{l|}{\cellcolor[HTML]{FFFFFF}{0.77}} &
  \multicolumn{1}{l|}{\cellcolor[HTML]{FFFFFF}{0.76}} &
  \multicolumn{1}{l|}{\cellcolor[HTML]{FFFFFF}{0.75}} &
  \multicolumn{1}{l|}{\cellcolor[HTML]{FFFFFF}{0.76}} &
  {0.12} &
  \multicolumn{1}{l|}{\cellcolor[HTML]{FFFFFF}{0.74}} &
  \multicolumn{1}{l|}{\cellcolor[HTML]{FFFFFF}{0.73}} &
  \multicolumn{1}{l|}{\cellcolor[HTML]{FFFFFF}{0.71}} &
  \multicolumn{1}{l|}{\cellcolor[HTML]{FFFFFF}{0.74}} &
  {0.14} \\ \cline{2-17} 
\rowcolor[HTML]{FFFFFF} 
\multicolumn{1}{|c|}{\cellcolor[HTML]{FFFFFF}} &
  {word2vec + tf–idf + RF} &
  \multicolumn{1}{l|}{\cellcolor[HTML]{FFFFFF}{0.80}} &
  \multicolumn{1}{l|}{\cellcolor[HTML]{FFFFFF}{0.52}} &
  \multicolumn{1}{l|}{\cellcolor[HTML]{FFFFFF}{0.53}} &
  \multicolumn{1}{l|}{\cellcolor[HTML]{FFFFFF}{0.63}} &
  {0.17} &
  \multicolumn{1}{l|}{\cellcolor[HTML]{FFFFFF}{0.79}} &
  \multicolumn{1}{l|}{\cellcolor[HTML]{FFFFFF}{0.74}} &
  \multicolumn{1}{l|}{\cellcolor[HTML]{FFFFFF}{0.75}} &
  \multicolumn{1}{l|}{\cellcolor[HTML]{FFFFFF}{0.77}} &
  {0.12} &
  \multicolumn{1}{l|}{\cellcolor[HTML]{FFFFFF}{0.76}} &
  \multicolumn{1}{l|}{\cellcolor[HTML]{FFFFFF}{0.62}} &
  \multicolumn{1}{l|}{\cellcolor[HTML]{FFFFFF}{0.64}} &
  \multicolumn{1}{l|}{\cellcolor[HTML]{FFFFFF}{0.68}} &
  {0.15} \\ \cline{2-17} 
\rowcolor[HTML]{FFFFFF} 
\multicolumn{1}{|c|}{\multirow{-12}{*}{\cellcolor[HTML]{FFFFFF}{\textbf{\rotatebox[origin=c]{90}{Classifier Chains}}}}} &
  {word2vec + tf–idf + XGB} &
  \multicolumn{1}{l|}{\cellcolor[HTML]{FFFFFF}{0.68}} &
  \multicolumn{1}{l|}{\cellcolor[HTML]{FFFFFF}{0.63}} &
  \multicolumn{1}{l|}{\cellcolor[HTML]{FFFFFF}{0.61}} &
  \multicolumn{1}{l|}{\cellcolor[HTML]{FFFFFF}{0.65}} &
  {0.18} &
  \multicolumn{1}{l|}{\cellcolor[HTML]{FFFFFF}{0.79}} &
  \multicolumn{1}{l|}{\cellcolor[HTML]{FFFFFF}{0.77}} &
  \multicolumn{1}{l|}{\cellcolor[HTML]{FFFFFF}{0.78}} &
  \multicolumn{1}{l|}{\cellcolor[HTML]{FFFFFF}{0.78}} &
  {0.11} &
  \multicolumn{1}{l|}{\cellcolor[HTML]{FFFFFF}{0.75}} &
  \multicolumn{1}{l|}{\cellcolor[HTML]{FFFFFF}{0.71}} &
  \multicolumn{1}{l|}{\cellcolor[HTML]{FFFFFF}{0.71}} &
  \multicolumn{1}{l|}{\cellcolor[HTML]{FFFFFF}{0.73}} &
  {0.14} \\ \hline
\rowcolor[HTML]{FFFFFF} 
\multicolumn{2}{|c|}{\cellcolor[HTML]{FFFFFF}{\textbf{BERT Classifier}}} &
  \multicolumn{1}{l|}{\cellcolor[HTML]{FFFFFF}{\color[HTML]{212121} \textbf{\underline{0.86}}}} &
  \multicolumn{1}{l|}{\cellcolor[HTML]{FFFFFF}{\color[HTML]{212121} \textbf{\underline{0.86}}}} &
  \multicolumn{1}{l|}{\cellcolor[HTML]{FFFFFF}{\color[HTML]{212121} \textbf{\underline{0.86}}}} &
  \multicolumn{1}{l|}{\cellcolor[HTML]{FFFFFF}{\color[HTML]{212121} \textbf{\underline{0.86}}}} &
  {\color[HTML]{212121} \textbf{\underline{0.14}}} &
  \multicolumn{1}{l|}{\cellcolor[HTML]{FFFFFF}0.88} &
  \multicolumn{1}{l|}{\cellcolor[HTML]{FFFFFF}0.88} &
  \multicolumn{1}{l|}{\cellcolor[HTML]{FFFFFF}0.88} &
  \multicolumn{1}{l|}{\cellcolor[HTML]{FFFFFF}0.88} &
  {\color[HTML]{212121} 0.13} &
  \multicolumn{1}{l|}{\cellcolor[HTML]{FFFFFF}{\color[HTML]{212121} 0.83}} &
  \multicolumn{1}{l|}{\cellcolor[HTML]{FFFFFF}{\color[HTML]{212121} 0.83}} &
  \multicolumn{1}{l|}{\cellcolor[HTML]{FFFFFF}{\color[HTML]{212121} 0.83}} &
  \multicolumn{1}{l|}{\cellcolor[HTML]{FFFFFF}{\color[HTML]{212121} 0.83}} &
  {\color[HTML]{212121} 0.17} \\ \hline
  \rowcolor[HTML]{FFFFFF} 
\multicolumn{2}{|c|}{\cellcolor[HTML]{FFFFFF}{\textbf{RoBERTa Classifier}}} &
  \multicolumn{1}{l|}{\cellcolor[HTML]{FFFFFF}{\color[HTML]{212121} 0.77}} &
  \multicolumn{1}{l|}{\cellcolor[HTML]{FFFFFF}{\color[HTML]{212121} 0.77}} &
  \multicolumn{1}{l|}{\cellcolor[HTML]{FFFFFF}{\color[HTML]{212121} 0.77}} &
  \multicolumn{1}{l|}{\cellcolor[HTML]{FFFFFF}{\color[HTML]{212121} 0.77}} &
  {\color[HTML]{212121} 0.23} &
  \multicolumn{1}{l|}{\cellcolor[HTML]{FFFFFF}0.88} &
  \multicolumn{1}{l|}{\cellcolor[HTML]{FFFFFF}0.88} &
  \multicolumn{1}{l|}{\cellcolor[HTML]{FFFFFF}0.88} &
  \multicolumn{1}{l|}{\cellcolor[HTML]{FFFFFF}0.88} &
  {\color[HTML]{212121} 0.13} &
  \multicolumn{1}{l|}{\cellcolor[HTML]{FFFFFF}{\color[HTML]{212121} \textbf{\underline{0.88}}}} &
  \multicolumn{1}{l|}{\cellcolor[HTML]{FFFFFF}{\color[HTML]{212121} \textbf{\underline{0.88}}}} &
  \multicolumn{1}{l|}{\cellcolor[HTML]{FFFFFF}{\color[HTML]{212121} \textbf{\underline{0.88}}}} &
  \multicolumn{1}{l|}{\cellcolor[HTML]{FFFFFF}{\color[HTML]{212121} \textbf{\underline{0.88}}}} &
  {\color[HTML]{212121} \textbf{\underline{0.13}}} \\ \hline
  \rowcolor[HTML]{FFFFFF} 
\multicolumn{2}{|c|}{\cellcolor[HTML]{FFFFFF}{\textbf{DistilBERT Classifier}}} &
  \multicolumn{1}{l|}{\cellcolor[HTML]{FFFFFF}{\color[HTML]{212121} 0.79}} &
  \multicolumn{1}{l|}{\cellcolor[HTML]{FFFFFF}{\color[HTML]{212121} 0.79}} &
  \multicolumn{1}{l|}{\cellcolor[HTML]{FFFFFF}{\color[HTML]{212121} 0.79}} &
  \multicolumn{1}{l|}{\cellcolor[HTML]{FFFFFF}{\color[HTML]{212121} 0.79}} &
  {\color[HTML]{212121} 0.21} &
  \multicolumn{1}{l|}{\cellcolor[HTML]{FFFFFF}\textbf{\underline{0.89}}} &
  \multicolumn{1}{l|}{\cellcolor[HTML]{FFFFFF}\textbf{\underline{0.89}}} &
  \multicolumn{1}{l|}{\cellcolor[HTML]{FFFFFF}\textbf{\underline{0.89}}} &
  \multicolumn{1}{l|}{\cellcolor[HTML]{FFFFFF}\textbf{\underline{0.89}}} &
  {\color[HTML]{212121} \textbf{\underline{0.11}}} &
  \multicolumn{1}{l|}{\cellcolor[HTML]{FFFFFF}{\color[HTML]{212121} \textbf{\underline{0.88}}}} &
  \multicolumn{1}{l|}{\cellcolor[HTML]{FFFFFF}{\color[HTML]{212121} \textbf{\underline{0.88}}}} &
  \multicolumn{1}{l|}{\cellcolor[HTML]{FFFFFF}{\color[HTML]{212121} \textbf{\underline{0.88}}}} &
  \multicolumn{1}{l|}{\cellcolor[HTML]{FFFFFF}{\color[HTML]{212121} \textbf{\underline{0.88}}}} &
  {\color[HTML]{212121} \textbf{\underline{0.13}}} \\ \hline
  
\end{tabular}}
\end{table*}

\subsubsection{Deep Learning}
We used BERT {\cite{devlin2018bert}, RoBERTa \cite{liu2019roberta}, and DistilBERT \cite{sanh2019distilbert}}, which {are the} state of the art classifiers designed to bi-directionally train on the overall context of data in all layers. BERT is pre-trained on Wikipedia and Book corpus that is helpful in English text. However, since our dataset may have words that are technical and unknown to the model \cite{bhargava2021generalization}, we processed and trained the model on pre-trained data and by adding more output layers. {Robustly optimised BERT approach (RoBERTa) is a replication study of BERT pretraining that measures the impact of key hyperparameters and training data size on top of BERT. RoBERTa is trained with dynamic masking, FULL-SENTENCES without Next Sentence Prediction (NSP) loss, large mini-batches and a larger byte-level Byte-Pair Encoding (BPE) \cite{liu2019roberta}. DistilBERT is another extension of BERT that is pre-trained on a smaller general-purpose language representation model and can then be fine-tuned on a wide range of tasks resulting in good performances. DistilBERT is reported to reduce the size of a BERT model by 40\%, while retaining 97\% of its language understanding capabilities and being 60\% faster, and also cheaper to pre-train \cite{sanh2019distilbert}.
}

As the first step, we needed to transform the data as required for these three models. Hence, we followed data pre-processing that allows them to interpret the data.

\textbf{Pre-processing.}
For BERT, RoBERTa and DistilBERT classifiers, we followed the same steps as described in Section \ref{sec:ML}. Additionally, 
we needed to transform the text data into numerical values within the word limit specified by Transformer models (512 words at a time). 
Therefore, we tokenised and encoded the data in a structure as specified by HuggingFace (transformers) library \cite{Huggingface2021}. The package provides API to perform NLP tasks to utilise its capacity over a vast variety of tasks \cite{prathibhamol2016multi}.

\textbf{{Classification Models.}}
The BERT classifier is ready to be utilised directly on the model, but it needs to be fine-tuned to produce better results. In a text classification task, the BERT base model outputs a vector of length 768 for each word (token) and pools the output. The pooled output contains overhead information while training, which helps in improving the predictions. {Given we achieved sufficiently useful results for these three models, in order to keep the models comparable, we used the same settings as BERT for RoBERTa and DistilBERT models.}


\subsubsection{Performance Metrics}
We considered 75\%-25\% ratio for our training and test sets, and used different metrics to measure the performance of our models.

\textit{Accuracy:}
Accuracy provides the ratio of correctly labelled data to complete data as either \emph{right} or \emph{wrong} \cite{mathews2021ah}. This disregards the possibility of partial correctness of labels, and hence the disadvantage of this measure is that multi-label classification problems may report low score for models capability to identify some labels from the set \cite{sorower2010literature}.

\textit{Micro precision and recall:}
Precision and Recall can be calculated in either Micro or Macro averages. We rely on Micro for our evaluation as it considers the aggregate contribution of all classes for the metric.

\textit{Micro F1:}
Micro F1 is the harmonic mean of micro precision and micro recall which accounts for both false positives and false negatives. It takes frequency of label as a contributor into consideration while evaluating performance of a model. Hence, micro averaged F1 score works extremely well in highly imbalanced distribution of tags \cite{medium2019detailed}. 

\textit{Hamming loss:}
In a multi label classification problem, partially correct predictions are not rewarded and hence the model performance cannot be determined completely. Hamming loss accounts for partially correct prediction to the total label set and hence it measures each label individually \cite{prathibhamol2016multi}. It reports the average number of times that an example is  incorrectly predicted to be related to a class label. It considers the prediction error, i.e. predicting an incorrect label, and missing error, i.e. not predicting a relevant label, normalised over total number of classes and examples \cite{medium2022valuation}.
Hamming Loss is interpreted from top bottom, with lower value as better score. 

\subsection{Findings}
We present the highest scoring results from different ngram combinations 
of ML in addition to our DL methods in Table \ref{tbl:MLResults}. We applied multiple models and feature extraction strategies for a multi-label approach, and hence, there is a collection of each classifier model that is paired to every method. 
The results are reported based on our two datasets and their combination. 

Results from the App review dataset show that app reviews are notably more random and hard for models to learn. Among ML models, we observed that using [Classifier chains + TF-IDF (analyser = char, ngram = 4,4)] for feature extraction along with employing Logistic Regression, we were able to achieve a relatively high score in most of the performance metrics (Accuracy-0.66, F1 score-0.7278 and Hamming Loss-0.148). Nonetheless, among our DL models, the BERT classifier, performed significantly better than the highest scoring ML model by a marginal difference from its baseline performance (Accuracy 0.8583, F1 Score 0.8583 and Hamming Loss 0.141). 

In contrast, models performed significantly better on the Issue Comments dataset as the language of comments was professional and involved more technical terminology. Thus, highest score of a ML model is  with a combination of [Classifier chain + TF-IDF (analyser = char, ngram = 4,4)] with base Linear SVM performance (Accuracy-0.786, F1 score-0.797, and Hamming loss-0.104). {Overall, DistilBERT has the highest overall performance among all ML/DL methods (Accuracy-0.885, F1 score-0.885, and Hamming loss-0.114).}

By combining the datasets to see if increasing the comment count (training size) will improve the performance, model performances were drastically reduced by an average of 3-5\% { in ML models}. Individual ML model performance is not high in all metrics and hence it becomes aimless to consider them. {RoBERTa and DistilBert take the lead with higher scores in all metrics and equal Hamming Loss as some other methods. }
Overall, among the machine learning methods, the linear SVM model and Logistic regression outperformed in different settings (different datasets) in our test run. We also noted higher performance with the Classifier chain as ML Strategy. Similarly, one best feature extraction technique is TF-IDF with ‘char’ combinations of ngram set as (4,4) that was observed to deliver the highest scores. We have considered multiple metrics but we prioritise the F1 score as our primary score due to the highly unbalanced nature of our datasets which makes it ideal. {As our results indicate, the BERT, RoBERTa and DistilBERT classifiers significantly outperformed the other models for different datasets in multi label text classification for highly imbalanced data by a good margin.} 

\begin{center}
\begin{tcolorbox}[colback=white!2!white,colframe=black!75!black]
\textbf{RQ2 Summary.} ML and DL algorithms can be used to automatically detect end-user human-centric issues from  developer  discussions and app reviews. {The best results were achieved using BERT on app reviews, DistilBERT on issue comments, and both RoBERTa and DistilBERT classifiers on the combination of the two datasets.}
\end{tcolorbox}
\end{center}

\section{Perceived Usefulness of Automatically Classifying End-user Human-centric Issues (RQ3)}\label{sec:RQ3appfindings}
\subsection{Approach} {We performed a survey to investigate the usefulness of the automated classification of human-centric issues from the perspective of software/app developers.}

\textbf{Protocol.} 
{We used the guidelines proposed by Kitchenham and Pfleeger \cite{kitchenham2008personal} to design and execute our survey. The survey introduction included three parts: \textit{Problem Statement}, \textit{Approach}, and \textit{Survey Goal}. In the \textit{Problem Statement} part, we defined the human-centric issue concept and depicted our taxonomy of human-centric issues. As the goal of the survey was not to ask the practitioners to use the automated approaches (discussed in Section \ref{sec:RQ2appfindings}) in practice before providing the feedback, we introduced the functionality of these approaches to the survey respondents in the \textit{Approach} part. To this end, we showed 8 user app reviews from the Signal Private app that BERT, as one of the best-performing automated approaches, could correctly classify them into Non-Human-centric Issues or Human-centric Issues. We also indicated that the approach classified those reviews labelled as Human-centric Issues into one or more of the following categories of human-centric issues: App Usage, Inclusiveness, and User Reaction. The next item (\textit{Survey Goal}) was to describe the objectives of the survey. The survey was anonymous and hosted on the Qualtrics platform. We obtained ethics approval from the Human Ethics Committee at Monash University before initiating the research.}

Our survey included 11 questions and took 5-7 minutes to complete. All questions except one were compulsory. Out of 11 questions, 5 questions sought the background information of the participants (e.g., ``\textit{what is your main role in software development?}''). The survey had 6 Likert scale questions and one optional open-ended question to allow the participants to share additional comments on the survey. The Likert scale questions asked the participants to rate the extent they agree or disagree with 6 statements (from ``strongly agree = 5'' to ``strongly disagree = 1''). We also added the ``I Don't Know'' option to the Likert questions to not force the respondents to answer the statements that they were unsure about or were unclear to them. We leveraged the survey studies \cite{nasab2021automated, palacio2019learning} used to evaluate the usefulness of (the outputs of) ML/DL-based approaches and tools in the software engineering community to design the following statements. 

\begin{table*}[]
\centering
\caption{Demographics information of the survey respondents}
\label{tbl:Tablesurvey}
{\scriptsize
\begin{tabular}{lllll}
\hline
\textbf{Country} & \textbf{Experience} & \textbf{Main role} & \textbf{Organisation size}                & \textbf{Organisation domain} \\ \hline
China     & 0-2 years  & DevOps Engineer   & More than 1000 employees                  & Others                     \\
Angola    & 6-10 years & Developer         & 100 \textless employees \textless{}= 500  & E-commerce                 \\
Australia & 3-5 years  & Software Engineer & Less than 20 employees                    & Consulting and IT services \\
India     & 0-2 years  & Project Manager   & Less than 20 employees                    & Consulting and IT services \\
India     & 0-2 years  & UI/UX Designer    & More than 1000 employees                  & Others                     \\
India            & 3-5 years           & Project Manager    & 20 \textless{}= employees \textless{}= 50 & Consulting and IT services   \\
Australia        & 0-2 years           & Business Analysis  & 20 \textless{}= employees \textless{}= 50 & Consulting and IT services   \\
India     & 0-2 years  & UI/UX Designer    & Less than 20 employees                    & Healthcare                 \\
Australia        & 3-5 years           & UI/UX Designer     & 100 \textless employees \textless{}= 500  & Consulting and IT services   \\
India     & 3-5 years  & Developer         & 20 \textless{}= employees \textless{}= 50 & E-commerce                 \\
India     & 0-2 years  & Consultant        & Less than 20 employees                    & Consulting and IT services \\
India     & 0-2 years  & Software Engineer & More than 1000 employees                  & Consulting and IT services \\
India     & 0-2 years  & Consultant        & More than 1000 employees                  & Consulting and IT services \\
India     & 0-2 years  & Developer         & More than 1000 employees                  & Consulting and IT services \\
India     & 0-2 years  & Developer         & Less than 20 employees                    & Consulting and IT services \\ 
    Canada     & 11-15 years  &  Software Engineer & 500 \textless employees \textless{}= 1000                    & Financial \\
\hline
\end{tabular}}
\end{table*}

\begin{itemize}

\item {Statement 1. ``\textit{The tool is useful because issue comments or app reviews with human-centric issues identified by the tool convey meaningful and important information}''.}

\item {Statement 2. ``\textit{The tool is useful because issue comments or app reviews with human-centric issues identified by the tool can be used to make informed human-centric issues-related design decisions in the future or refine the existing sub-optimum decisions}''.}

\item {Statement 3. ``\textit{The tool is useful because I, as a practitioner, can find meaningful and important information in a reasonable timeframe from issue comments or app reviews with human-centric issues identified by the tool}''.}

\item {Statement 4. ``\textit{The tool is useful because issue comments or app reviews with human-centric issues identified by the tool can help us identify human-centric issues faster in mobile apps than if we did it manually}''.}

\item {Statement 5. ``\textit{The tool is useful because issue comments or app reviews with human-centric issues identified by the tool may contain information that can help us prioritize and resolve such human-centric issues in mobile apps more effectively}''.}

\item {Statement 6. ``\textit{The tool is useful because issue comments or app reviews with human-centric issues identified by the tool can provide hints and cues to trace forward and backward to codes, services, or features that lead to human-centric issues}''.}

\end{itemize}

\textbf{Participants.} We recruited the survey participants by broadly advertising our survey on social networks like Twitter and LinkedIn. In total, we got 16 valid responses. Table \ref{tbl:Tablesurvey} outlines the demographic information of the survey respondents including the country they currently work in, the total number of years they have been involved in software development, their main role in software development, the size of their organisation, and the main domain of their organisation. Most of the respondents were working in India (10) followed by Australia (3), Canada (1), China (1) and Angola (1). They mostly had less than two years (10), followed by 3-5 years (4), 6-10 years (1) and 11-15 years (1) of experience in software development. We had four developers, three UI/UX Designers, three software engineers, two project managers, two consultants, one DevOps engineer, and one business analyst among the respondents. Five reported their organisation to have more than 1000, one between 500 and 1000, two to have between 100 and 500, three between 20 and 50, and five with less than 20 employees. Finally, most organisations (10) were in Consulting and IT services domain, two in E-commerce, one in healthcare, one in Finance and two other domains.

\textbf{Data Analysis.} We applied descriptive statistics to analyse the closed-ended questions, i.e., demographic and Likert scale questions.
\subsection{Findings}

{\renewcommand{\arraystretch}{1.5}
\begin{table}[]
\centering
\caption{Participants' level of agreement with the statements (in \%).}\label{tbl:surveyresponses}
\scalebox{0.84}{
\begin{tabular}{l c c c c c c } 
\Xhline{3\arrayrulewidth}
              & \textbf{\begin{tabular}[c]{@{}c@{}}Strongly \\ Agree\end{tabular}} & \textbf{Agree} & \textbf{Neutral} & \textbf{Disagree} & \textbf{\begin{tabular}[c]{@{}c@{}}Strongly \\ Disagree \end{tabular}} & \textbf{\begin{tabular}[c]{@{}c@{}}I Don't \\ Know\end{tabular}} \\ \Xhline{3\arrayrulewidth}
\textbf{Statement 1}            & 
18.75        &
\cellcolor[HTML]{7CFF54} 50       & 
31.25        & 
0.0      & 
0.0         & 
0.0              \\
\textbf{Statement 2}                                                                                         &    
 25        & 
  \cellcolor[HTML]{7CFF54} 43.75        & 
31.25        & 
 0.0        & 
 0.0         & 
 0.0          \\
\textbf{Statement 3}  & 
18.75      & 
\cellcolor[HTML]{7CFF54} 68.75       & 
12.5       & 
 0.0        & 
 0.0        & 
 0.0       \\
\textbf{Statement 4}                                                                                                                                 & 
\cellcolor[HTML]{7CFF54} 37.5        & 
37.5       & 
25       & 
 0.0         & 
 0.0         & 
 0.0               \\
\textbf{Statement 5}                                                                                                        &            
 25       & 
\cellcolor[HTML]{7CFF54} 62.5      & 
12.5       & 
 0.0        & 
 0.0        & 
 0.0

\\
\textbf{Statement 6}                                                                                       &                        
25       & 
31.25       & 
 \cellcolor[HTML]{7CFF54} 43.75      & 
 0.0        & 
 0.0        & 
 0.0                \\ \Xhline{3\arrayrulewidth}
\end{tabular}}
\end{table}}
Participants' level of agreement to Statements 1-6 is shown in Table \ref{tbl:surveyresponses}. Among all the respondents, the majority strongly agreed (18.75\%) or agreed (50\%) that the tool is useful since the human-centric issues include noteworthy information (Statement 1). Regarding the usefulness of the tool to help make better human-centric issues-related design decisions or refine the existing ones, 25\% strongly agreed and 43.75\% agreed (Statement 2). 18.75\% of the participants strongly agreed, and 68.75\% agreed that the tool can help find meaningful and important information in a reasonable timeframe from the human-centric issues detected by the tool (Statement 3). 75\% of the respondents strongly agreed/agreed with Statement 4, indicating the human-centric issues detected by the tool can help practitioners identify human-centric issues faster in mobile apps than if they did it manually (Statement 4). 

Among the survey respondents, the vast majority of the respondents (87.5\%) strongly agreed/agreed with the usefulness of the tool as the human-centric issues detected by the tool contain information that can help them prioritise and resolve such issues in mobile apps more effectively (Statement 5). Finally, 25\% of the respondents strongly agreed and 31.5\% agreed that the detected issues can provide suggestions to tace to codes, services, or features that lead to human-centric issues (Statement 6). None of the respondents disagreed/strongly disagreed with any of the statements. One developer mentioned that \textit{``These applications are vital to identify issues beforehand, which can be used to avoid such issues any further.''}

\begin{center}
\begin{tcolorbox}[colback=white!2!white,colframe=black!75!black]
\textbf{RQ3 Summary.} Practitioners indicate that the automated classification of end-user human-centric issues discussed in issue comments and mobile app reviews has several applications in practice, particularly in enabling practitioners to find meaningful and important information related to human-centric issues in a reasonable timeframe and helping practitioners prioritise and resolve human-centric issues more effectively.
\end{tcolorbox}
\end{center}

\section{Discussion}\label{sec:Discussion}

\textit{\textbf{There are discrepancies between human-centric issues reported by the users and discussed by the developers;}}
In our data set we found there are almost twice as many human-centric issues reported in the sample 1,200 app reviews (47.25\%) compared to the sample 1,200 issue comments (25.5\%). 
{Having a lower proportion of human-centric issues discussed in issue comments was foreseeable given the issue comments discuss the possible solution that might be reported through several app reviews}. In the App Usage category, both users/developers most frequently mention UI/UX and Bugines issues, followed by users discussing  changes/updates, while developers more frequently discuss Privacy/security. Users report many app access issues, but we did not find any discussions between developers to resolve such issues. We found several app reviews mentioning monetary aspects (we classified as Others), but no discussions of these in developers comments. 
In the Inclusiveness category, developers discuss many language, compatibility and accessibility related issues. Interestingly, such issues are rarely reported by the app users. This could be because developers have already discussed and resolved such issues and therefore, users did not experience many challenges. Finally, in the User Reaction category, users mainly report Preference and Fulfilling interests related issues. While developers also more often discuss Preference issues, they rarely discuss Fulfilling interests issues. 
These discrepancies can indicate that users face issues that are not being discussed by the developers, and at the same time, if developers discuss and resolve human-centric issues, users do not experience such issues. Therefore, developers need to be aware of the human-centric issues that the users report, and carefully discuss and resolve them during app development.

\textit{\textbf{Human-centric issues are different across projects;}}
Our findings show that the prevalence of human-centric issues varies across different projects, as illustrated in Tables \ref{tbl:TableIssue} and \ref{tbl:TableApp}. In line with the previous discussion point, all the studied projects have less human-centric issues discussed via issue comments than reported by the users. There are projects with both frequent (e.g. Signal private messenger
) and rare (e.g. Pixel-dungeon game
) human-centric issues discussed by developers and reported by users. In some apps, Firefox browser as the most obvious example, users have very frequently reported human-centric issues (79 out of 100), noting this is just the number of comments containing at least one human-centric issue and some of them may include issues from several categories. However, there are very limited human-centric issues discussed by the developers (30 out of 100). Another examples are K-9 Mail 
and Bitcoin, the app for buying and using bitcoin and crypto. 
On the other hand, some apps have relatively frequent discussions of human-centric issues in issue comments but not in app reviews. Cgeo, an app for geocaching, has relatively high occurrence of human-centric issues in issue comments (33 out of 100), but not that of app reviews (29 out of 100). Similarly, Wordpress, the website building app, has 26 human-centric issue comments and only 28 app reviews. The rest of the apps follow more or less the same general trend, and have almost twice app reviews comparing to issue comments. This encourages future research to study how human-centric issues are impacted in different projects or devise guidelines for designing human-centric apps for general users apps (such as Firefox, K-9, Bitcoin).

{\textit{\textbf{Human-centric issues can be both technical and non-technical;}}
Some of the categories, such as buginess, UI/UX, and compatibility might be perceived as technical issues, rather than human-centric issues. However, we have only considered an issue, including a technical issue, as a human-centric issue if it directly impacts the end-users of the app. We did not want to limit ourselves to any specific issues or ignore any kind of issues experienced by the end-users of the app if they are being explored as technical issues by other researchers if we believed they were related to the end users' human characteristics. The aim was to be able to build a comprehensive taxonomy that stems from the challenges that the end-users face. This is aligned with what is considered in the work by Ramos et al. \cite{ramos2021considerations}. Ramos et al. present a scoping review of 68 studies that employed one or more assessment tools to evaluate a mental health app. They aim to identify the extent to which the existing app evaluation frameworks capture diversity, equity, and inclusion factors. This paper has a different objective from ours, however, has various categories of diversity, equity, and inclusion to assess the app evaluation frameworks. Their diversity, equity, and inclusion criteria were adapted from the Culturally-Informed Design Framework \cite{valdez2012designing}. Ramos et al. provide definitions of extracted diversity, equity, and inclusion variables in three domains: \textit{access}, \textit{content}, and \textit{appearance}. Similar to our categories, their \textit{access} domain covers variables such as Internet Connectivity, Data Usage, Cost, and System Requirements, which seem to be technical but are considered to measure Diversity, Equity, and Inclusion. Considering a comprehensive set of issues faced by the end-users led to the identification of the “Inclusiveness” category in our work that has not been explored in the past. This category can be the focus of other researchers’ future work.
}

\textit{\textbf{There is no structured way of reporting and addressing most human-centric issues;}}
We found that there is currently no structured way for the users to report human-centric issues through app reviews, also found in other work on defect reporting \cite{yusop2016reporting}. Moreover, human-centric issues are mostly discussed from a technical perspective by the developers on GitHub issue comments. This suggests a need for a more human-centric issue reporting and follow-up process and tools. Issue reporting systems should include relevant details from not only a technical perspective but also a non-technical end-user understandable point of view. Our future work would allow users to report different human-centric issues, and incorporate such issues in a systematic way during the app development process. 
{There exist some preliminary works on human-centric defect reporting. A recent work by Huynh et al. \cite{huynh2021improving} captured a subset of specific disabilities, e.g., colour blindness, dyslexia, aphasia, hearing impairment, dexterity impairment, and vision impairment, using personas, and further developed mobile and web application prototypes to support defect reporting for this diverse user background. While the work of Huynh et al. \cite{huynh2021improving} provides some preliminary defects reporting support for these end-users, it does not capture our broader objective of reporting a wider range of human-centric issues. Potential guidelines and features for such a tool would include a relatively simple in-app form(s), user tutorials on how to effectively report human-centric defects, form sections for how to reproduce human-centric defects (expected vs actual results), form sections to categorise the type of human-centric issues encountered, and user-reported level of criticality, amongst others.}
Such reporting tools should, of course, themselves be human-centric and support a diverse range of end-users of the reporting tools. There should also be better ways for the end-users and developers to communicate and become aware of human-centric issues. {Our future work would pursue this line of research.}

\textit{\textbf{Awareness of human-centric issues can help developers and researchers to more effectively incorporate and report  human-centric issues;}}
Limited discussion about key human-centric issues among developers, in spite of the frequent reporting of such issues, reflects that there is still an ongoing challenge that lies in front of the software industry to design more human-centric software and mobile apps. 
Developers need to be more aware of the human-centric issues of their end-users in order to design more inclusive and human-centred apps and to avoid negative impacts on different end-user groups. Software engineers are typically very different from most end-users - a profession heavily dominated by men; relatively young; affluent; technical; most proficient in English; and while some have physical/mental challenges, these are generally different or of less severity than many users, especially for software targeted to challenged end-users \cite{grundy2020towards,grundy2020humanise}. These influence the degree that developers appreciate and know how to address human-centric issues of their end-users. Training the developers, supporting them by providing required resources, and increasing their general awareness of the human-centric issues could improve the consideration of these issues during the development process. Results from \cite{Alshayban:2020} indicate the importance of accessibility awareness to make app developers becoming ambassadors of accessibility in their organisations.

\textit{\textbf{An automated tool may help detect human-centric issues from app reviews and GitHub issue comments;}} Our ML/DL models performance and feedback we collected through surveying practitioners suggest that an automated way of detecting human-centric issues can be useful for both users and app developers. {Automatically categorising and prioritising app reviews can help developers in different ways. Some examples include suggesting the maintenance tasks developers have to accomplish by extracting the topics and classifying the intention of the reviewer \cite{di2016would}, ranking informative reviews in order to support app developers in identifying and prioritising numerous informative reviews \cite{malgaonkar2022prioritizing}, and identifying common patterns in order to detect performance bugs for smartphone applications and further support follow-up research on avoiding performance bugs, testing, debugging and analysis for mobile phone applications \cite{liu2014characterizing}. Our work would help not only developers but also users to be able to project their issues and challenges to the developers through an automated tool that is able to detect such issues.} Developers can more easily compile such issues among a huge pile of reviews they receive on a regular basis and make sure they are aware and can account for the issues reported by the users. Moreover, developers can easily search through the GitHub issue comments to understand what human-centric issues are already discussed in a project when contributing to a new project. Future research is needed to analyse other software (e.g. Jira, StackOverflow) and user-base repositories (e.g. user stories) and apply such smart tools to different sources of data. Furthermore, this can encourage other researchers and practitioners to set some actionable items and guidelines to incorporate human aspects in different software development stages to avoid the occurrence of these issues.

\section{Threats to Validity}\label{sec:ThreatsValidity}
This section discusses possible threats or limitations of this study and the approaches adopted to mitigate threats \cite{wohlin2012experimentation}.

\subsection{Internal Validity}
\textbf{Data Collection.} The selection of the 12 studied projects, 1200 issue comments, and 1200 app reviews from each project may have introduced threats to our study. First, we decided to use Mazuera-Rozo et al. \cite{mazuera2020investigating}'s dataset, consisting of 100 Android projects randomly collected from an extensive dataset of GitHub Android projects. We then applied a set of criteria to reach 12 Android projects. This decision was motivated by the fact that it was not possible for us to manually analyse the issue comments and app reviews of all 100 Android projects. Although the 12 projects have different characteristics (e.g., they come from diverse domains), we accept there might be some important Android projects from the human-centric issue perspective that have been omitted. Second, manual analysis of all issue comments and app reviews from these 12 projects was not feasible. Hence, we randomly selected a subset of issue comments and app reviews from each project. We may have missed important developer/end-user discussions on human-centric issues from these projects.

\textbf{Manual Classification (RQ1).}  The qualitative analysis used to build the taxonomy of human-centric issues might be subjective and error-prone. To this end, three analysts were involved in the qualitative analysis, and the taxonomy was built in two phases. In each phase, each issue comment/app review was independently analysed and labelled by two persons. Any disagreements between two analysts on labelling issue comments were resolved either by open discussions or involving the third analyst in the discussions. To avoid possible risks and mistakes, when it was not clear to identify the type of human-centric issue from a given issue comment, we labelled it as a non-human-centric issue. Hence, we are confident that our taxonomy of human-centric issues is credible with minimum mistakes.


\textbf{Survey (RQ3).} Completing the survey did not require particular skills from  the practitioners. They only had to have some level of software/app development experience. Still, the respondents with poor knowledge of software/app development could threaten the validity of the survey's findings. To partially mitigate this threat, the ``I Don't Know'' option was added to each Likert scale question.

\subsection{External Validity} 

\textbf{Manual Classification (RQ1).} Two factors can threaten the generalisability of the findings of \textbf{RQ1}. Firstly, the 12 selected projects are a small subset of all Android projects hosted on GitHub. Furthermore, our dataset does not include any iOS projects. 
We acknowledge that our taxonomy of human-centric issues may not be generalised to all different types of GitHub projects (e.g., iOS projects). Secondly, the identified categories of human-centric issues are exclusive to developer discussions on GitHub and end-user reviews on the Google Play Store and are not comprehensive. Hence, analysing other open-source software repositories (e.g., Bitbucket) and software artefacts (e.g., commits, requirement specifications) of proprietary and open-source projects may lead to identifying different and/or a more comprehensive set of human-centric issues categories.


\textbf{Survey (RQ3).} The number of responses (16 responses) to our survey is comparable to other similar surveys (e.g., \cite{prana2019categorizing, abualhaija2020automated}) used to investigate the usefulness of ML/DL approaches in software engineering. Despite this, the survey findings cannot be generalised to all software/mobile app practitioners, software/mobile app development organisations, and types of apps. Further to this, the majority of the respondents came from India, which is another threat to external validity.

\subsection{Construct Validity}

\textbf{Experiments (RQ2).} We used several feature extractions techniques  with four ML classifiers and developed one DL classifier (See Section \ref{sec:RQ2appfindings}). We also used four metrics, as precision, recall, accuracy, F1-score, and hamming loss to evaluate ML/DL classifiers. Many other feature selection techniques, ML/DL classifiers, and metrics could be used. However, as argued by Peters et al. \cite{peters2017text}, it was impossible for us to implement all ML/DL classifiers and use all metrics in one study. The used feature selection techniques, DL/ML classifiers, and metrics are widely used in automating and classifying software engineering tasks. Despite this, we confirm that the use of other future selection techniques, ML/DL classifiers, and metrics can lead to different results. Finally, there are many data splitting methods \cite{xu2018splitting}. We decided to use 75\%-25\% to split our dataset into the training and test set in our study. We acknowledge that using other splitting methods may lead to different performance results.

\textbf{Survey (RQ3).} The concept of \textit{human-centric issues} may have different meanings for practitioners. In the survey introduction, we explicitly defined human-centric issues to avoid misunderstanding or multi interpretation of human-centric issues among the survey respondents. 
{In this study, we used a survey to evaluate the usefulness of our automated learning approaches (e.g., BERT) in detecting and classifying human-centric issues from the perspective of software/app developers. While surveys are a common approach to show the usefulness of ML/DL approaches in software engineering, we should emphasise that our survey respondents did not use the developed automated approaches. They only assess the usefulness of the approaches based on their functionality introduced in the survey, along with some examples of app reviews that were correctly detected and classified by BERT as one of the approaches. We are aware that the real drawbacks and practical merits of an ML/DL approach can only be revealed when practitioners use it in practice. Hence, other types of research methodologies, for example, industrial case studies and user studies need to be conducted to explore all drawbacks and practical merits of our ML/DL approaches in practice.}

\subsection{Reliability}
Another threat that might impact this study is that others attempt to replicate it but achieve different results. We constructed a replication package \cite{anonymous_2021_4739069} and made it publicly available for those who want to replicate, validate, or extend this study. The replication package includes the dataset used to build the taxonomy, the results of ML and DL approaches, and raw responses to the survey.

\section{Related Work}\label{sec:RelatedWork}


Online repositories, question and answer sites, and issue tracking platforms such as GitHub, StackOverflow, and Jira not only contain rich data discussing technical aspects of the software development process but also include information that provides insight into the \textbf{social and human aspects} of the software development process \cite{Ortu:2016}. GitHub has been of considerable interest to software engineering researchers for years \cite{kalliamvakou2014promises} due to many open source projects and rich technical and non-technical information to be mined. Many of the projects hosted on GitHub are public, and therefore anyone can view the activities, including actions around issues, pull requests, and commits within those projects. 

\subsection{Human Aspects in Software Development}

Addressing the role of technical proficiency in the software development process, Rocetti et al. compared two approaches in participatory design of a large software artefact involving: 1) novice users, and 2) expert users. Their results show that most of the innovative proposals came from novice users \cite{Roccetti:2020}. This shows that designing human-centric software artefacts requires a more participation from novice users, in contrast to the traditional opinion that expert users provide more reliable contribution to the software design process. Rauf et al. analysed a dataset of app developers to examine the rationale behind developers’ prioritisation of security in the software development process \cite{Rauf:2020}. The study shows that social considerations, e.g., fear of users, influenced developers’ reasoning in development activities, including security choices \cite{Rauf:2020}.

Moreover, a human aspect that has been discussed in recent years is the concept of human values (i.e., the guiding principles of what people consider important in life \cite{Cheng:2010}), and its relationship with technology \cite{Grundy:2022address, Hussain:2020}. Whittle et al. argued for the consideration and inclusion of human values at different stages in the software development life cycle \cite{Whittle:2021}. Another study introduced a set of interventions for addressing human values in the SAFe Agile framework \cite{Hussain:2022How}.  Other related works have proposed tools for supporting human values, e.g., a human values dashboard for software development \cite{Arif:2021towards}, values Q-sort - an instrument for capturing the values of software engineers \cite{winter2018measuring}, and algorithms for detection the violation of values in Android APIs \cite{li2021step}.
{\subsection{Human Aspects in Software Repositories}}

Pletea et al. focused on security-related discussions on GitHub, as mined from discussions around commits and pull requests \cite{pletea2014security}. 
Ko et al. analysed developer design discussions through Bugzilla bug reports to understand the design challenges and how the decisions are made to adapt to user needs \cite{ko2011design}. Twidale et al. focused on usability bug reports in Bugzilla \cite{twidale2005exploring} while Andreasen et al. explored developers' opinions about usability through surveys, interviews, and mining software repositories \cite{andreasen2006usability}. Studies have also mined social aspects in repositories. Dabbish et al. mined GitHub for transparency and collaboration in GitHub projects \cite{Dabbish:2012}, while Dam et al. mined open-source projects for social norms \cite{Dam:2015}. Barcellini et al. analysed and visualised social, thematic temporal, and design aspects of online software repositories to understand and model the dynamics of the open source software design process in mailing list exchanges \cite{barcellini2008socio}. 

Some works have focused on mining and classifying specific human aspects of developers in software repositories and issue tracking platforms. Mining more than 2 million issues in Jira from 4 open-source software projects \cite{Ortu:2015}, Ortu et al. found a positive correlation between developers’ emotions and issue fixing time. Positive emotions resulted in shorter issue-fixing time while negative emotions related to longer issue-fixing time. Cabrera-Diego et al. developed classifiers for comments related to emotions on StackOverflow and Jira. Using features derived from different lexica, their results show significant improvements over the current state of the art in emotion classification \cite{CABRERADIEGO:2020}. Another study analysed software artefacts for the presence of emotional information in the software development process \cite{Murgia:2014}. Results of an analysis of the Apache Software Foundation issue tracking show that developers do express emotion while discussing technical issues. Although these studies focus on a specific human aspect (i.e., emotion) from a developer's perspective, they indicate that a rational view of the software development process is insufficient; human aspects such as emotions can negatively or positively affect the development process and be propagated into the resulting software artefact, e.g., happiness, a positive emotion, increases creativity \cite{Fredrickson2001}, which is good for a successful software design \cite{Brooks:1987}. 

Khalajzadeh et al. \cite{khalajzadeh2022diverse} conducted an empirical study of issue comments by extracting and manually analysing 1,691 issue comments from 12 diverse projects, ranging from small to large-scale projects. They categorised the human-centric issues into eight categories of: Inclusiveness, Privacy \& Security, Compatibility, Location \& Language, Preference, Satisfaction, Emotional Aspects, and Accessibility. However, this work only focuses on developers viewpoint on human-centric issues through analysing GitHub issue comments. 

\begin{table*}[]
\centering
\caption{Comparison with other works on categorising app reviews (Y = Yes, N = No, P = Partial)}
\label{tbl:appreviewcategories}
{\scriptsize
\renewcommand{\arraystretch}{1.5}

\begin{tabular}{|l|lllllll|llll|lll|}
\hline
 &
  \multicolumn{7}{c|}{\textbf{\#1: App Usage}} &
  \multicolumn{4}{c|}{\textbf{\#1: Inclusiveness}} &
  \multicolumn{3}{c|}{\textbf{\#3: User Reaction}} \\ \hline
Category &
  \multicolumn{1}{l|}{\textbf{\rotatebox[origin=c]{90}{Resource Usage}}} &
  \multicolumn{1}{l|}{\textbf{\rotatebox[origin=c]{90}{Buginess}}} &
  \multicolumn{1}{l|}{\textbf{\rotatebox[origin=c]{90}{Change/Update}}} &
  \multicolumn{1}{l|}{\textbf{\rotatebox[origin=c]{90}{UI/UX}}} &
  \multicolumn{1}{l|}{\textbf{\rotatebox[origin=c]{90}{Privacy/security}}} &
  \multicolumn{1}{l|}{\textbf{\rotatebox[origin=c]{90}{Usage instruction}}} &
  \textbf{\rotatebox[origin=c]{90}{Access issues}} &
  \multicolumn{1}{l|}{\textbf{\rotatebox[origin=c]{90}{Compatibility}}} &
  \multicolumn{1}{l|}{\textbf{\rotatebox[origin=c]{90}{Location}}} &
  \multicolumn{1}{l|}{\textbf{\rotatebox[origin=c]{90}{Language}}} &
  \textbf{\rotatebox[origin=c]{90}{Accessibility}} &
  \multicolumn{1}{l|}{\textbf{\rotatebox[origin=c]{90}{Fulfilling interests}}} &
  \multicolumn{1}{l|}{\textbf{\rotatebox[origin=c]{90}{Emotional aspects}}} &
  \textbf{\rotatebox[origin=c]{90}{Preference}} \\ \hline
Khalid et al. \cite{khalid2014mobile} &
  \multicolumn{1}{l|}{Y} &
  \multicolumn{1}{l|}{Y} &
  \multicolumn{1}{l|}{N} &
  \multicolumn{1}{l|}{P} &
  \multicolumn{1}{l|}{Y} &
  \multicolumn{1}{l|}{N} &
  N &
  \multicolumn{1}{l|}{Y} &
  \multicolumn{1}{l|}{N} &
  \multicolumn{1}{l|}{N} &
  N &
  \multicolumn{1}{l|}{N} &
  \multicolumn{1}{l|}{N} &
  P \\ \hline
McIlroy et al.\cite{mcilroy2016analyzing} &
  \multicolumn{1}{l|}{Y} &
  \multicolumn{1}{l|}{Y} &
  \multicolumn{1}{l|}{Y} &
  \multicolumn{1}{l|}{P} &
  \multicolumn{1}{l|}{Y} &
  \multicolumn{1}{l|}{N} &
  N &
  \multicolumn{1}{l|}{Y} &
  \multicolumn{1}{l|}{N} &
  \multicolumn{1}{l|}{N} &
  N &
  \multicolumn{1}{l|}{N} &
  \multicolumn{1}{l|}{N} &
  P \\ \hline
Chen et al. \cite{chen2021should} &
  \multicolumn{1}{l|}{N} &
  \multicolumn{1}{l|}{N} &
  \multicolumn{1}{l|}{N} &
  \multicolumn{1}{l|}{P} &
  \multicolumn{1}{l|}{N} &
  \multicolumn{1}{l|}{N} &
  N &
  \multicolumn{1}{l|}{N} &
  \multicolumn{1}{l|}{N} &
  \multicolumn{1}{l|}{N} &
  Y &
  \multicolumn{1}{l|}{N} &
  \multicolumn{1}{l|}{N} &
  N \\ \hline
Sorbo et al. \cite{di2016would} &
  \multicolumn{1}{l|}{Y} &
  \multicolumn{1}{l|}{Y} &
  \multicolumn{1}{l|}{Y} &
  \multicolumn{1}{l|}{P} &
  \multicolumn{1}{l|}{Y} &
  \multicolumn{1}{l|}{N} &
  P &
  \multicolumn{1}{l|}{P} &
  \multicolumn{1}{l|}{N} &
  \multicolumn{1}{l|}{N} &
  N &
  \multicolumn{1}{l|}{N} &
  \multicolumn{1}{l|}{N} &
  P \\ \hline
Alshayban et al. \cite{Alshayban:2020} &
  \multicolumn{1}{l|}{N} &
  \multicolumn{1}{l|}{N} &
  \multicolumn{1}{l|}{N} &
  \multicolumn{1}{l|}{N} &
  \multicolumn{1}{l|}{N} &
  \multicolumn{1}{l|}{N} &
  N &
  \multicolumn{1}{l|}{N} &
  \multicolumn{1}{l|}{N} &
  \multicolumn{1}{l|}{N} &
  Y &
  \multicolumn{1}{l|}{N} &
  \multicolumn{1}{l|}{N} &
  N \\ \hline
Obie et al. \cite{Obie:2021} &
  \multicolumn{1}{l|}{N} &
  \multicolumn{1}{l|}{N} &
  \multicolumn{1}{l|}{N} &
  \multicolumn{1}{l|}{N} &
  \multicolumn{1}{l|}{Y} &
  \multicolumn{1}{l|}{N} &
  N &
  \multicolumn{1}{l|}{N} &
  \multicolumn{1}{l|}{N} &
  \multicolumn{1}{l|}{N} &
  N &
  \multicolumn{1}{l|}{Y} &
  \multicolumn{1}{l|}{N} &
  Y \\ \hline
Fazzini et al. \cite{fazzini2022characterizing} &
  \multicolumn{1}{l|}{N} &
  \multicolumn{1}{l|}{N} &
  \multicolumn{1}{l|}{N} &
  \multicolumn{1}{l|}{P} &
  \multicolumn{1}{l|}{Y} &
  \multicolumn{1}{l|}{N} &
  N &
  \multicolumn{1}{l|}{P} &
  \multicolumn{1}{l|}{Y} &
  \multicolumn{1}{l|}{Y} &
  Y &
  \multicolumn{1}{l|}{N} &
  \multicolumn{1}{l|}{Y} &
  P \\ \hline
\end{tabular}}
\end{table*}

{\subsection{Human Aspects in App Reviews}}
Alshayban et al. conducted a large-scale study to understand the state of accessibility in android apps and found that accessibility issues are rife in the 1,000 apps they studied. In some cases, mobile app developers are not educated in accessibility principles and/or are not incentivised by their organisations to make their apps more accessible \cite{Alshayban:2020}. Furthermore, a recent study on the reflection of human values in mobile app reviews shows that a quarter of the 22,119 app reviews analysed contain perceived violation of human values in mobile apps, supporting the recommendation for the use of app reviews as a potential source for mining values requirements in software projects \cite{Obie:2021}. 

{There has been some studies on categorising app reviews from the users' point of view. Khalid et al. \cite{khalid2014mobile} studied user reviews from 20 iOS apps and uncovered 12 types of user complaints: App Crashing, Compatibility, Feature Removal, Feature Request, Functional Error, Hidden Cost, Interface Design, Network Problem, Privacy and Ethics, Resource Heavy, Uninteresting Content, Unresponsive App, and Not Specific.
McIlroy et al. \cite{mcilroy2016analyzing} studied reviews from 20 mobile apps in the Google Play Store and Apple App Store and proposed an approach to automatically assign multiple labels to app reviews. They categorised the app reviews in the following groups: Additional Cost, Functional Complaint, Compatibility Issue, Crashing, Feature Removal, Feature Request, Network Problem, Other, Privacy and Ethical Issue, Resource Heavy, Response Time, Uninteresting Content, Update Issue, User Interface.} 

{Chen et al. \cite{chen2021should} identified four categories of user interface related issues: Appearance, Interaction, Experience, and Other generic UI related issues by manually analysing a random sample of 1,447 reviews out of the 3.3M UI-related app reviews. They further categorised these four issue categories into 17 UI-related issue types that users are concerned about in reviews. The Appearance category includes Layout, Legibility and Colour, Typography and Font, Iconography, and Image. The Interaction category includes Navigation, Notification, Motion, Gesture, and Accessibility. The Experience category consists of subcategories as Redundancy, Customisation, Advertisement, and Feedback. Finally, the Others category covers Generic Review, Comparative Review, and Design Specification. All these categories and sub-categories focus on user interface-related issues. 
Sorbo et al. \cite{di2016would} introduce SURF (Summariser of User Reviews Feedback), which automatically extracts the topics in app reviews, and classifies the intention of the reviewer to suggest the maintenance tasks developers have to accomplish. They categorise the intentions as Information Giving, Information Seeking, Feature Request, Problem Discovery, and Other. They also group together sentences covering the same topic, such as App, GUI, Contents, Pricing, Feature or Functionality, Improvement, Updates/Versions, Resources, Security, Download, Model, and Company.}

{Genc-Nayebi et al. \cite{genc2017systematic} conducted a systematic literature review to identify the proposed solutions for mining app store user reviews, challenges and unsolved problems, contributions to software requirements evolution and future research directions in the domain. They provided a summary of the extracted app features in a list of mobile app feature extraction studies. Feature request, bug report, compatibility, customer support, updates, user experience, privacy, and resources are the features that are consistent with our findings. Fazzini et al. \cite{fazzini2022characterizing} conducted
an empirical study focusing on app reviews of COVID-19 contact
tracing apps. By manually analysing a dataset of 2,611 app reviews, they categorised them into nine categories of Age, Disability, Emotion, Gender, Language, Location, Privacy, Socioeconomic, and Miscellaneous. Even though this work has focused on human aspects, it is limited to the inclusiveness related aspects of the applications, specifically in COVID-19 contact tracing apps. 
Table \ref{tbl:appreviewcategories} shows how our categories in this paper are covered in the summarised studies.}

{\subsection{Summary}}

All of the studies discussed in this section focus on different human and social aspects and provide insight into how these aspects are represented in the software development process and repositories. 
However, none of these works provide an analysis of how human-centric aspects of the end-users are discussed by both end-users and developers in the same projects. 
In addition, there currently does not exist a comprehensive taxonomy of human-centric issues from both developers and end-users point of view. 
Our work fills this important gap by providing a broader view perspective of these discussions, with a focus on \textbf{end-user human-centric issues}. 
This is the first work to look into these human-centric issues from both end-users and developers perspective, and also propose an automated way to detect such issues and validate it with real practitioners. In this paper, we developed categories for these human aspects based on a manual analysis of issue comments from different software projects on GitHub and app reviews of the same projects on Google Play Store. 


\section{Conclusion}\label{sec:Conclusion}
Based on a manual analysis of 2,400 app reviews and issue comments from 12 different GitHub repositories, we investigated 
what human-centric issues are raised by the end-users on Google Play Store and discussed by developers of the same projects on GitHub. We categorised the human-centric issues reported by end-users on Google Play Store app reviews, and discussed by developers in GitHub issue comments into three high-level categories: App Usage, Inclusiveness, and User Reaction. We reflected on the fact that there is no standard way of reporting and addressing such human-centric issues on both Google Play Store and GitHub repositories. We also developed ML/DL models to help developers with very different human aspects to many of their end-users to be able to automatically detect such human-centric issues. The results of our ML/DL models in addition to the feedback we received from 16 software/app practitioners supported that our approach can help developers to recognise and appreciate such diverse software end-user human-centric issues more easily. In our future work, we plan to investigate other repositories, question and answer sites, and issue tracking platforms, such as Jira and Stack Overflow. {We believe there is a lot of space in further exploring the new “Inclusiveness” category that has emerged from our analysis. It can be the focus of our or other researchers’ future work, and our automated tool can be used on larger scale datasets extracted from other repositories to be able to detect the issues related to the inclusiveness category and further explore its sub-categories.} We also plan to formulate human-centric requirements to be able to model and incorporate them in different software development stages. 


\section*{Acknowledgment}
Support for this work from ARC Laureate Program FL190100035 and ARC Discovery DP200100020 is gratefully acknowledged.



%

\bibliographystyle{IEEEtran}
\bibliography{References}

\begin{IEEEbiography}[{\includegraphics[width=1in,height=1.25in,clip,keepaspectratio]{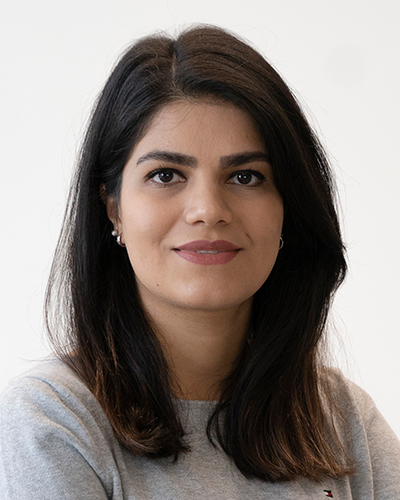}}]{Hourieh Khalajzadeh} is a Senior Lecturer in the School of IT at Deakin University. She was previously a Research Fellow in the HumaniSE Lab at Monash University. Her research interests include human-centred software engineering and model-driven software engineering. She has received several awards such as ACM Distinguished Paper Award (MobileSoft 2022), Outstanding Reviewer Award (CHASE 2021), Best Paper Award (ENASE 2020), and Best Showpiece Award (VLHCC 2020). 

\end{IEEEbiography}

\begin{IEEEbiography}[{\includegraphics[width=1in,height=1.25in,clip,keepaspectratio]{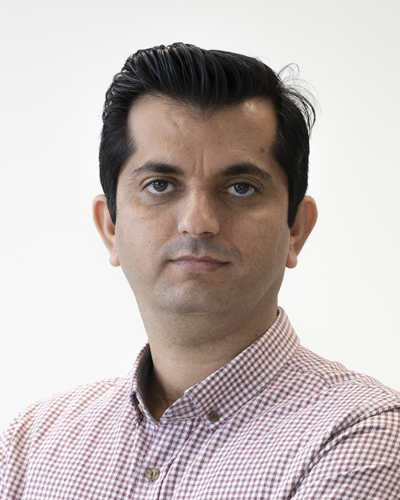}}]{Mojtaba Shahin} is a Lecturer in the School of Computing Technologies at RMIT University, Melbourne. Previously, he was a Research Fellow at Monash University. His research interests reside in Empirical Software Engineering, Human and Social Aspects of Software Engineering, and Secure Software Engineering. He completed his PhD study at the University of Adelaide, Australia.
\end{IEEEbiography}

\begin{IEEEbiography}[{\includegraphics[width=1in,height=1.25in,clip,keepaspectratio]{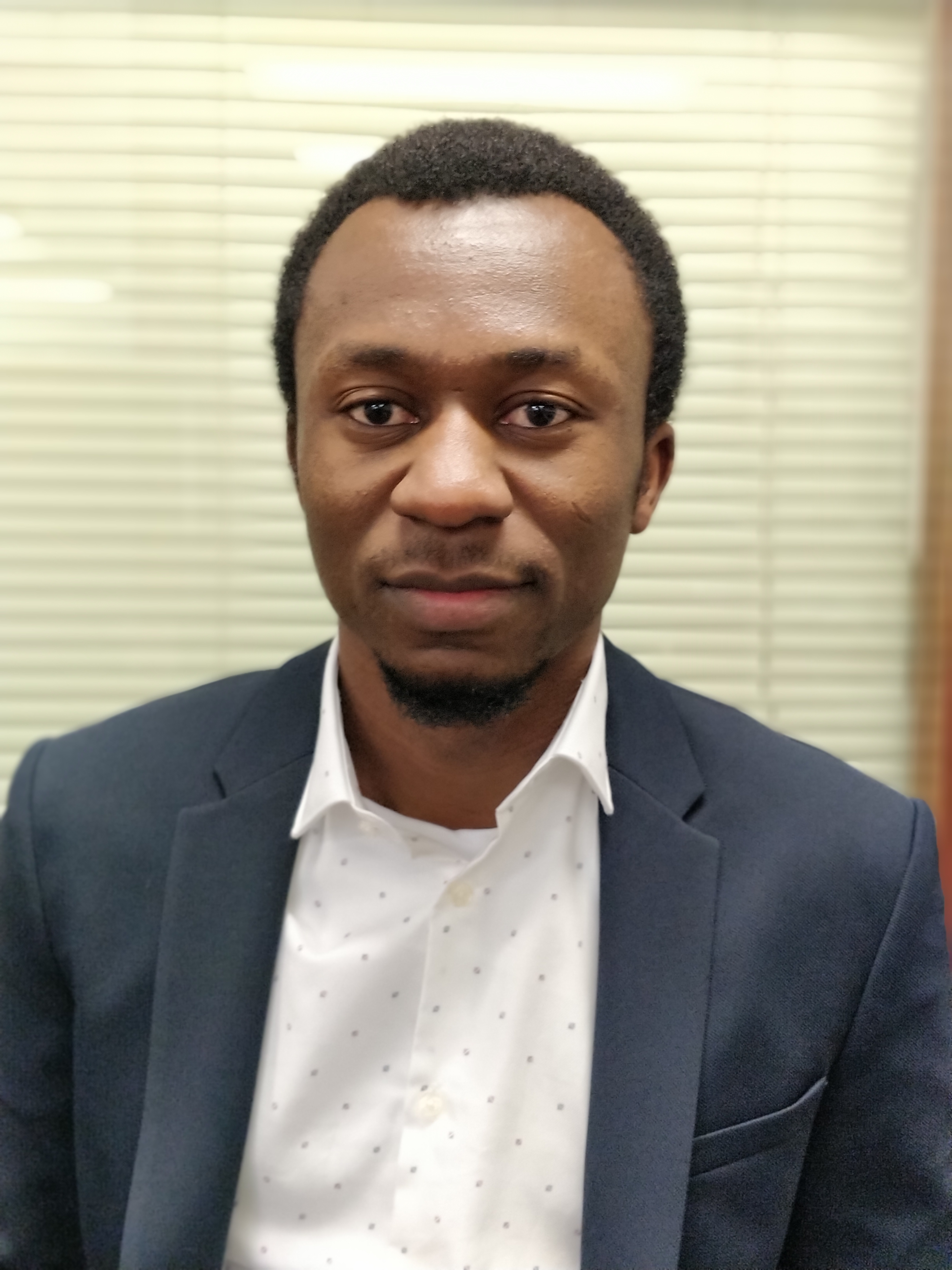}}]{Humphrey O. Obie} is an Adjunct Research Fellow with the HumaniSE Lab in the Faculty of Information Technology, Monash University. He has tackled problems in several domains as a data scientist, software developer, and researcher. His key interests include human-centric software engineering, human-centric IoT and smart cities, value-based software engineering, information visualisation and visual data storytelling. He completed his PhD at Swinburne University of Technology, Australia.
\end{IEEEbiography}

\begin{IEEEbiography}[{\includegraphics[width=1in,height=1.25in,clip,keepaspectratio]{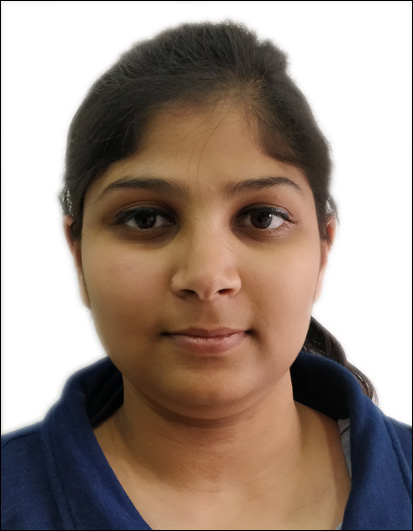}}]{Pragya Agrawal} is a second-year Master of Data Science student at Monash University, Australia. She has received a bachelor’s degree in Computer and Communication Engineering from Manipal University Jaipur, India. Her current research interests include Data mining and Statistical Analysis.
\end{IEEEbiography}

\begin{IEEEbiography}[{\includegraphics[width=1in,height=1.25in,clip,keepaspectratio]{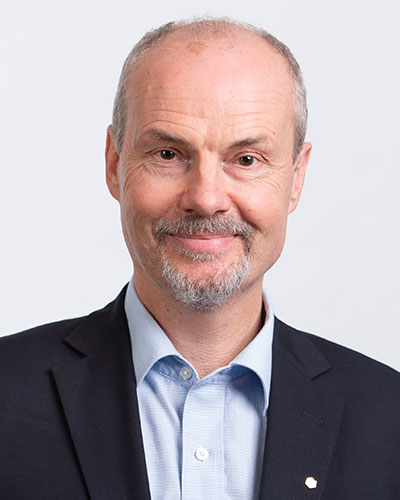}}]{John Grundy} is an Australian Laureate Fellow and a Professor of Software Engineering at Monash University where he heads up the HumaniSE lab in the Faculty of Information Technology.  His lab investigates “human-centric” issues in software engineering – these include, but are not limited to, impact of personality on software engineers and users; emotion-oriented requirements engineering and acceptance testing; impact of different languages, cultures and belief sets on using software and engineering software; usability and accessibility of software, particularly for ageing people and people with physical and mental challenges; issues of gender, age, socio-economic status and personal values on software, software requirements, and software engineering; and team and organisational impacts, including team climate. His lab's goal is to improve software engineering practices, tools, and thus the target systems to make better software for people.
\end{IEEEbiography}

\end{document}